%% file: main.tex
\documentclass[12pt]{article}
\usepackage{amsmath}
\usepackage{amsfonts}
\usepackage{graphicx}
\usepackage{enumerate}
\usepackage{mathtools}
\usepackage{natbib}
\usepackage{url} 
\usepackage{amsthm}
\usepackage{bm}
\usepackage{booktabs}
\usepackage{verbatim}
\usepackage{color}
\newcommand{\blind}{1}

\newtheorem{theorem}{Theorem}[section]
\newtheorem{lemma}[theorem]{Lemma}
\newtheorem{definition}[theorem]{Definition}
\newtheorem{corollary}[theorem]{Corollary}
\newtheorem{proposition}{Proposition}
\newtheorem{remark}{Remark}
\newcommand{\Var}{\mathrm{Var}}
\newcommand{\Cov}{\mathrm{Cov}}

\newcommand{\augx}{[ \bm X  \bm{\widetilde X} ]}
\newcommand{\augbeta}{[\bm{\widehat\beta}^\top \bm{\widetilde\beta}^\top]}
\newcommand{\augOATK}{[\bm{X}_{\backslash j} \; \bm{\widetilde x}_j]}

\DeclareMathOperator{\sign}{sign}
\DeclareMathOperator{\diag}{diag}
\DeclareMathOperator{\SSE}{SSE}
\newcommand{\RR}{\mathbb{R}}
\newcommand{\EE}{\mathbb{E}}
\newcommand{\PP}{\mathbb{P}}

\newcommand{\ind}{\mathbf{1}}
\newcommand{\fdr}{\textnormal{FDR}}
\newcommand{\fdp}{\textnormal{FDP}}

\newcommand{\cH}{\mathcal{H}}
\newcommand{\cS}{\mathcal{S}}
\newcommand{\eps}{\varepsilon}
\newcommand{\covp}{\stackrel{\textnormal{p.}}{\rightarrow}}
\newcommand{\afdp}{\overline{\fdp}}
\newcommand{\indep}{\perp\!\!\!\!\perp} 
\newcommand{\bX}{{\bm X}}
\newcommand{\bS}{{\bm S}}

\newcommand{\si}{\textnormal{SI}}
\renewcommand{\hat}{\widehat}
\newcommand{\bU}{\bm{U}}
\mathtoolsset{showonlyrefs}
\begin{document}

\def\spacingset#1{\renewcommand{\baselinestretch}%
{#1}\small\normalsize} \spacingset{1}


\if1\blind
{
  \title{\bf One-at-a-time knockoffs: controlled false discovery rate with higher power}
  \author{Charlie K. Guan, Zhimei Ren, Daniel W. Apley}
  \maketitle
} \fi

\if0\blind
{
  \bigskip
  \bigskip
  \bigskip
  \begin{center}
    {\LARGE\bf One-at-a-time knockoffs: controlled false discovery rate with higher power}
\end{center}
  \medskip
} \fi

\bigskip
\begin{abstract}
We propose one-at-a-time knockoffs (OATK), a new methodology for detecting important explanatory variables in linear regression models while controlling the false discovery rate (FDR). For each explanatory variable, OATK generates a knockoff design matrix that preserves the Gram matrix by replacing one-at-a-time only the single corresponding column of the original design matrix. OATK is a substantial relaxation and simplification of the knockoff filter by \cite{BC_2015}(BC), which simultaneously generates all columns of the knockoff design matrix to satisfy a much larger set of constraints.
To test each variable's importance,  statistics are then constructed by comparing the original vs. knockoff coefficients.
Under a mild correlation assumption on the original design matrix, OATK asymptotically controls the FDR at any desired level. Moreover, OATK consistently achieves (often substantially) higher power than BC and other approaches across a variety of simulation examples and a real genetics dataset. Generating knockoffs one-at-a-time also has substantial computational advantages and facilitates additional enhancements, such as conditional calibration or derandomization, to further improve power and consistency of FDR control. OATK can be viewed as the conditional randomization test (CRT) generalized to 
fixed-design linear regression problems, and can generate fine-grained p-values 
for each hypothesis. 

\end{abstract}

\noindent%
{\it Keywords:}  variable selection, knockoff filter, selective inference, false discovery rate, multiple hypothesis testing
\vfill

\newpage
\spacingset{1.9} 
\section{Introduction}
\label{sec:intro}

Linear and ridge regression are universal tools for data analysis. In applications in which there exist many explanatory variables in the data that do not affect the response variable, users often seek to discover the sparse subset of significant variables and discard the variables having no effect. Consider the fixed-design linear regression model 
\begin{equation}
    \label{eq:linear_model}
    \bm{y} = \bm{X\beta} + \bm{z}
\end{equation}
where $\bm{y}\in\mathbb{R}^n$ is the response vector, $\bm{X}=[\bm{x_1}, \bm{x_2}, \hdots, \bm{x_p}]\in\mathbb{R}^{n\times p}$ is the known and deterministic design matrix of explanatory variables with Gram matrix $\bm{\Sigma}=\bm{X}^\top\bm{X}$, $\bm{z}\sim N_p(0, \sigma^2 \bm{I}_p)$ is Gaussian noise, and $\bm{\beta}=[\beta_1, \beta_2, \hdots, \beta_p]^\top$ is the vector of unknown regression coefficients. 
Given this model, we are interested in testing the null hypotheses 
\begin{align*}
H_j: \beta_j=0 \quad  
\end{align*} 
for each $j\in[p] := \{1,2\ldots,p\}$. The goal is to identify $S=\{j:\beta_j\neq0\}\subseteq [p]$, i.e., the index set of non-null variables. 
Specifically, we aim to leverage the data to produce an estimate 
$\widehat{S}\subseteq [p]$ of  $S$, while controlling 
the false discovery rate $\text{FDR} = \mathbb{E}\left[ \text{FDP} \right]$, where 
\begin{equation}
    \label{eq:fdp_fdr} \text{FDP} = \frac{\#\{j:\beta_j=0 \text{ and } j\in \widehat{S}\}}
     {\#\{j: j\in \widehat{S}\} \vee 1} 
\end{equation}
is the false discovery proportion, and $a \vee b = \max(a,b)$ for $a,b \in \RR$.
Given a pre-specified target FDR level $\alpha \in (0,1)$, 
our primary goal is to control the FDR of $\widehat S$ by $\alpha$ in a manner that yields high $\text{Power} = \mathbb{E}\left[ \text{TDP} \right]$, where 
\begin{equation}
    \label{eq:tdp_power} \text{TDP} = \frac{\#\{j\in S \text{ and } j\in \widehat{S}\}}
     {\#\{j: j\in S\}} 
\end{equation}
is the true discovery proportion. In both \eqref{eq:fdp_fdr} and \eqref{eq:tdp_power}, the expectation is taken with respect to $\bm{z}$ with $\bm{X}$ treated as fixed. 

A classical approach to this variable selection problem is to first construct a p-value for 
each hypothesis, and then obtain a selection set based on these 
p-values~\citep{benjamini1995controlling,benjamini2001control,storey2002direct,storey2004strong}. 
In some cases, it might be difficult to obtain valid p-values, especially when one wishes 
to incorporate additional information (e.g., sparsity, smoothness) in the model fitting step.

Recently, there have been two new (suites of) p-value-free variable selection algorithms: 
the {\em knockoff filter}~\citep{BC_2015,candes2018panning} (BC)
and the {\em Gaussian mirror}~\citep{GM,dai2023false,dai2023scale} (GM). 
The knockoff filter constructs a knockoff matrix 
$\bm{\widetilde X} = [\bm{\widetilde x_1}, \bm{\widetilde x_2}, \hdots, \bm{\widetilde x_p}]\in\mathbb{R}^{n\times p}$ 
such that 
\begin{align*}
\bm{\widetilde{X}}^\top \bm{\widetilde{X}} = \bX^\top \bX,  \;\; \text{and}
\quad \bm{\widetilde X}^\top \bm{X} = \bm X^\top\bm{\widetilde{X}} = \bm{\Sigma}-\bm{S},
\end{align*}
where $\bm S = \text{diag}(\bm{s})$ and $\bm{s}$ is some suitably chosen $p$-dimensional non-negative vector, such that
$[\bX \bm{\widetilde X}]^\top [\bX \bm{\widetilde{X}}]$ is positive semi-definite.
The knockoff filter regresses (possibly using regularization, e.g., Lasso~\citep{tibshirani1996regression} or ridge~\citep{hoerl1970ridge}) 
$\bm y$ onto the augmented design matrix $[\bm X \bm{\widetilde X}]$ to yield the augmented vector of estimated coefficients $\augbeta = [\widehat\beta_1, \widehat\beta_2, \hdots, \widehat\beta_p, \widetilde\beta_1, \widetilde\beta_2, \hdots, \widetilde\beta_p ]$ for both the original explanatory variables and their knockoffs. The procedure then constructs certain antisymmetric statistics $\bm{W}=[W_1, \hdots, W_p]$ such that $W_j$ compares $\widehat\beta_j$ and $\widetilde\beta_j$ in a manner that results in a positive $W_j$ if $\widehat\beta_j$ appears more significant than $\widetilde\beta_j$ and a negative $W_j$ if the opposite is true. A variable is selected (i.e., $j\in\widehat S$) if $W_j$ is large and positive. For example, one choice that was further explored in~\citet{candes2018panning} is $W_j = |\widehat\beta_j|-\lvert\widetilde\beta_j\rvert$. Another choice that applies only to Lasso regression with regularization parameter $\lambda$, which was the primary focus of BC, is to define $W_j=\max(\widehat\lambda_j, \widetilde\lambda_j)\times \sign(\widehat\lambda_j-\widetilde\lambda_j)$, where $\widehat\lambda_j$ ($\widetilde\lambda_j$) denotes the largest $\lambda$ on the Lasso path for which $\widehat\beta_j$ ($\widetilde\beta_j$) was nonzero. A positive $W_j$ in this case means that, as $\lambda$ was increased, $\widehat\beta_j$ remained in the model longer than $\widetilde\beta_j$, and vice-versa for negative $W_j$.
The knockoff filter then selects variables via $\widehat S = \{j: W_j\geq T\}$, 
where the threshold $T$ is selected empirically via 
\begin{align}
    T\equiv\min \Bigg\{ t: \frac{1+\#\{j:W_j\leq -t\}}{\#\{j:W_j\geq t\} \vee 1} \leq \alpha \Bigg\}.
    \label{def:T_empirical}
\end{align}
Above, the numerator $1+\#\{j:W_j\leq -t\}$ is an estimate of the number of false discoveries for threshold $t$, 
since the knockoffs are constructed without regard for $\bm y$ and truly have no effect on the response, and so
$W_j$ is symmetric with respect to $0$ for the null features. \citet{BC_2015} showed that 
this knockoff procedure controls the FDR by $\alpha$.

Although the knockoff filter provably controls the FDR, it has been observed to lose power 
in certain problems~\citep{gimenez2019improving,liu2019power,li2021whiteout}.
The loss of power may be due to several reasons. Perhaps most significantly, the statistics $\bm W$ of the knockoff filter are computed for the augmented regression problem of regressing $\bm y$ onto $\augx$, which is substantially different than the original problem of regressing $\bm y$ onto $\bm X$. Namely, the augmented regression involves twice as many explanatory variables as the original regression, which typically substantially increases the variance of the estimated regression coefficients and reduces the ability to discern between $\beta_j=0$ versus $\beta_j\neq 0$. 

In contrast, the GM approach of \cite{GM} considers each explanatory variable one at a time. For each $\bm x_j$, 
it generates a pair of ``mirror variables'' $\bm x^+_j=\bm x_j+\bm z_j$ and $\bm x^-_j=\bm x_j-\bm z_j$ for some appropriately scaled Gaussian random vector $\bm z_j$, 
regresses $\bm y$ onto $[\bm{X}_{\backslash j}\; \bm{x}^+_j\; \bm{x}^-_j]$ where $\bm{X}_{\backslash j}$ denotes $\bm X$ with the $\bm x_j$ column removed,
and then computes a statistic $W_j$ by contrasting the coefficients for 
$\bm x^+_j$ and $\bm x^-_j$ in this augmented regression. As noted in \citet{GM}, this is equivalent to regressing $\bm y$ onto $[\bm X\; \bm z_j]$ and contrasting the coefficients of $\bm x_j$ and $\bm z_j$. 
Compared with knockoffs, this approach has weaker theoretical FDR control guarantees --- the FDR is controlled asymptotically 
when the dependence between features is mild --- but empirically, it sometimes exhibits
higher power while maintaining reasonable FDR control. 

We propose a new approach, termed {\em one-at-a-time knockoffs (OATK)}, 
that combines desirable aspects of the GM and BC approaches but that is couched more firmly in the knockoff framework. 
One of the contributions of this work is to draw a closer connection to the knockoff framework 
than was drawn in \cite{GM}, in a manner that provides insight into how to achieve more powerful detection of signals and improve computational issues, while preserving FDR control. 
The specific contributions of this work include:
\begin{itemize}
    \item [(1)] OATK constructs a knockoff copy for each variable one at a time, where the knockoffs must satisfy only a subset of the BC knockoff conditions, allowing for substantially greater flexibility in generating knockoffs. The subsequent knockoff regressions are of exactly the same dimension as the original regression (only a single column is replaced by its knockoff), avoiding the reduction in power associated with the BC regression onto $2p$ variables. Our feature importance statistics constructed from the knockoff and the original coefficients are still symmetric around zero for the nulls. 

    \item [(2)] OATK is far more computationally efficient to implement than the BC procedure, 
    and, using well-known matrix algebra identities, the feature importance statistics can be computed without actually regressing $\bm y$ onto 
    the knockoff design matrix. The low computational expense and one-at-a-time nature of the knockoffs facilitate adding performance enhancements such as derandomization or a ``multi-bit'' variant that constructs multiple knockoff copies for each 
    feature, leading to a fine-grained marginally valid p-value that can be viewed as a {\em conditional randomization test p-value}~\citep{candes2018panning}
    in the fixed-design setting. Such p-values (one-bit or multi-bit) are then passed to the 
    SeqStep filter~\citep{BC_2015} to generate the selection set.

    \item [(3)] We prove asymptotic FDR control of OATK under mild assumptions and develop non-asymptotic bounds on the FDR, characterizing conditions 
    for approximate FDR control. We also propose {\em conditionally-calibrated OATK} that achieves 
    exact FDR control, leveraging techniques in~\citet{fithian2022conditional}.
    
    \item [(4)] Through extensive numerical experiments and real-data examples, 
    we demonstrate that our approach is substantially more powerful than the approaches of either BC or GM, while still controlling the FDR.
\end{itemize}
 
\paragraph{Additional notation and assumptions used throughout the paper.} Throughout, we assume $\bm{X}$ is full rank, each $\bm{x}_j$ ($j\in [p]$) is scaled to have unit norm, and we denote $\bm{\widetilde X} = [\bm{\widetilde x}_1, \bm{\widetilde x}_2, \hdots, \bm{\widetilde x}_p]\in\mathbb{R}^{n\times p}$ as the knockoff matrix. The matrix $\bm{X}_{\backslash j}$ denotes the original design matrix $\bm X$ with column $j$ removed. For any matrix $\bm A$, let $\bm A_{\cdot j}$ and $\bm A_{j \cdot}$ denote the $j$-th column and $j$-th row, respectively, and let $A_{i j}$ or $[\bm A]_{ij}$ denote the $i$-th row, $j$-th column element. 
The set of null features is denoted by $\cH_0 = \{j: \beta_j = 0\}$. We denote $p_0 = |\cH_0|$ and $p_1=|S|$ as the number of null and non-null variables, respectively.

\section{One-at-a-time knockoff procedure}

Our approach generates knockoffs $\bm{\widetilde x}_j$ one-at-a-time satisfying the following two conditions:  
\begin{align}
    (a)~\bm{\widetilde X}^\top \bm X &= \bm \Sigma - \bm S, \quad
    \text{and}\quad 
    (b)~\bm{\widetilde x}_{j}^{\top} \bm{\widetilde x}_j = \bm x_j^\top\bm x_j=1
    \text{ for each }j\in[p],\label{eq:oat_cond} 
\end{align}
where $\bS = \text{diag}(\bm s)$ is a diagonal matrix with non-negative entries such that 
$\bm{\Sigma} - \bS$ is a positive semi-definite matrix. This is substantially less restrictive than the BC approach, which additionally requires $\bm{\widetilde X}^\top\bm{\widetilde X}=\bm\Sigma$. Throughout, the regressions can either be ordinary least squares (OLS) or regularized ridge regression, the former being a special case of the latter.

In Section \ref{subsec:structure}, we describe the main ideas behind our OATK regression. In Section \ref{subsec:generation}, we convert the knockoff requirements (\ref{eq:oat_cond}) into equivalent but computationally simpler conditions and suggest a simple and attractive choice for $\bm{S}$. In Section \ref{subsec:fastcomp}, we derive an explicit connection between our knockoff regression coefficients and the original regression coefficients that is both revealing and that allows a fast procedure for computing the coefficients without having to explicitly generate knockoffs or conduct the $p$ knockoff regressions. In Section \ref{subsec:w}, we discuss what properties the test statistics $\bm{W}$ must possess to control FDR. In Section \ref{sec:ext}, we describe other enhancements that OATK can incorporate to further improve FDR control and power.

\subsection{OATK regression structure}
\label{subsec:structure}
Consider the ridge regression parameter estimates
\begin{align}
    \label{eqn:ori_coef}
    \bm{\widehat\beta}_\lambda \equiv \begin{bsmallmatrix}
    \widehat\beta_{\lambda, 1} \\ \widehat\beta_{\lambda, 2} \\ \vdots \\ \widehat\beta_{\lambda, p}
\end{bsmallmatrix} = [\bm{X}^\top\bm{X} + \lambda \bm{I}_p]^{-1}\bm{X}^\top\bm{y} = \bm{\Sigma}_\lambda^{-1} \bm{X}^\top\bm{y},
\end{align}
where $\bm{\Sigma}_\lambda = \bm{ X}^\top \bm{ X} + \lambda \bm{I}_p$, and $\lambda$ is the ridge regularization parameter.  The same value for $\lambda$ used in this ridge regression is used in all the knockoff regressions described shortly. See Appendix \ref{appendix:fast_imp} for a computationally efficient procedure to select $\lambda$ empirically. We define our knockoff ridge regression approach as follows. For each $j\in[p]$, consider  
\begin{align}
    \bm{\widetilde\beta}^j_\lambda &\equiv \begin{bsmallmatrix}
        \widetilde\beta^j_{\lambda, 1} \\ \widetilde\beta^j_{\lambda, 2} \\ \vdots \\ \widetilde\beta^j_{\lambda, p}
    \end{bsmallmatrix} 
    = [(\bm{\widetilde X}^j)^{\top} \bm{\widetilde X}^j + \lambda \bm{I}_p]^{-1}(\bm{\widetilde X}^{j})^\top\bm{y} 
    = \bm{\widehat\beta}_\lambda + [\bm{\Sigma}^{-1}_\lambda]_{ 
    \cdot j}  
    \big(\bm{\widetilde{x}}_j^\top \bm y- \bm{x}_j^\top\bm y\big),
     \label{eqn:knock_reg}
\end{align}
where $\bm{\widetilde X}^j = \augOATK$ is the design matrix that we use in the OATK knockoff regression for the $j$-th variable (i.e., we replace only the $j$-th column by its knockoff). 
The only element of $\bm{\widetilde\beta}^j_\lambda$ that we will need is 
$\widetilde\beta_{\lambda, j}^j$, which from~\eqref{eqn:knock_reg} is obtained via 
$\widetilde\beta_{\lambda, j}^j = \widehat\beta_{\lambda, j} +[\bm{\Sigma}^{-1}_\lambda]_{jj}(\bm{\widetilde{x}}_j^\top\bm y - \bm{x}_j^\top\bm y)$, 
where $[\bm{\Sigma}^{-1}_\lambda]_{jj}$ can be interpreted as the reciprocal of the regression sum of squared error (SSE) for regressing $\bm{x}_j$ onto 
the other $p-1$ columns of $\bm X$. Thus, we can efficiently obtain our vector of knockoff coefficients via 
\begin{align}
    \label{eqn:beta_tilde_vec}
    \bm{\widetilde\beta}_\lambda &\equiv 
    \begin{bsmallmatrix}
        \widetilde\beta_{\lambda 1} \\ \widetilde\beta_{\lambda 2} \\ \vdots \\ \widetilde\beta_{\lambda p}
    \end{bsmallmatrix} \equiv 
    \begin{bsmallmatrix}
        \widetilde\beta_{\lambda 1}^1 \\ \widetilde\beta_{\lambda 2}^2 \\ \vdots \\ \widetilde\beta_{\lambda p}^p
    \end{bsmallmatrix} = 
        \bm{\widehat\beta}_\lambda + \begin{bsmallmatrix}
        [\bm{\Sigma}^{-1}_\lambda]_{11}(\bm{\widetilde{x}_1}^\top\bm y - \bm{x_1}^\top\bm y) \\
        [\bm{\Sigma}^{-1}_\lambda]_{22}(\bm{\widetilde{x}_2}^\top\bm y - \bm{x_2}^\top\bm y) \\
        \vdots \\
        [\bm{\Sigma}^{-1}_\lambda]_{pp}(\bm{\widetilde{x}_p}^\top\bm y - \bm{x_p}^\top\bm y)
    \end{bsmallmatrix}
\end{align}
Note that $\bm{\widehat\beta}_\lambda$ in \eqref{eqn:ori_coef} and \eqref{eqn:beta_tilde_vec} is different than in the augmented regression of the BC approach.

The following proposition, whose proof is in Appendix \ref{appendix:proof_distribution}, establishes the joint distribution of 
$(\bm{\widehat{\beta}}_\lambda, \bm{\widetilde{\beta}}_\lambda)$ with $\lambda$ treated as fixed.

\begin{proposition}
    \label{prop:distribution}
    For fixed $\lambda >0$, consider $\bm{\widehat{\beta}}_\lambda$ and $\bm{\widetilde{\beta}}_\lambda$ obtained from 
    Equations~\eqref{eqn:ori_coef} and~\eqref{eqn:beta_tilde_vec}, respectively. Then the joint distribution of $(\bm{\widehat{\beta}}_\lambda, \bm{\widetilde{\beta}}_\lambda)$ is
    \begin{align}
    \label{eqn:joint}
    \begin{bsmallmatrix}
        \bm{\widehat\beta}_\lambda \\ \bm{\widetilde\beta}_\lambda
    \end{bsmallmatrix} \sim \mathcal{N}_{2p}\bigg(\begin{bsmallmatrix}
        \bm{\mu_{\widehat \beta}} \\ \bm{\mu_{\widetilde \beta}}
    \end{bsmallmatrix}, \begin{bsmallmatrix}
        \bm{\Sigma_{\widehat \beta}} & \Cov^\top[\bm{\widetilde \beta}_\lambda, \bm{\widehat \beta}_\lambda] \\ \Cov[\bm{\widetilde \beta}_\lambda, \bm{\widehat \beta}_\lambda] & \bm{\Sigma_{\widetilde \beta}}
    \end{bsmallmatrix} \bigg),
    \end{align}
where 
\begin{align}
    &\bm{\mu_{\widehat\beta}} \equiv \bm{\Sigma}_\lambda^{-1}\bm{\Sigma\beta},\quad 
    \bm{\mu_{\widetilde \beta}}  \equiv
    (\bm{\Sigma}_\lambda^{-1}\bm\Sigma - \diag\{\bm{\Sigma}_\lambda^{-1}\}\bm S)\bm{\beta},\quad
    \bm{\Sigma_{\widehat\beta}} \equiv 
    \sigma^2\bm{\Sigma}_\lambda^{-1}\bm \Sigma \bm{\Sigma}_\lambda^{-1},\\
    &\bm{\Sigma_{\widetilde \beta}} \equiv 
    \sigma^2 [\bm{\Sigma}_\lambda^{-1}\bm \Sigma \bm{\Sigma}_\lambda^{-1} - \bm{\Sigma}_\lambda^{-1}\bm S \diag\{\bm{\Sigma}_\lambda^{-1}\} - \diag\{\bm{\Sigma}_\lambda^{-1}\}\bm S \bm{\Sigma}_\lambda^{-1} + \\
    &\qquad\qquad\qquad\qquad\qquad \diag\{\bm{\Sigma}_\lambda^{-1}\}(\bm{\widetilde X}^\top\bm{\widetilde X}-\bm\Sigma+2\bm S) \diag\{\bm{\Sigma}_\lambda^{-1}\}],\\
    &\Cov[\bm{\widetilde \beta}_\lambda, \bm{\widehat \beta}_\lambda] 
    = \sigma^2(\bm{\Sigma}_\lambda^{-1}\bm \Sigma \bm{\Sigma}_\lambda^{-1} - \diag\{\bm{\Sigma}_\lambda^{-1}\}\bm S\bm{\Sigma}_\lambda^{-1}).
\end{align}
\end{proposition}
From Proposition~\ref{prop:distribution},  
if $j\in \cH_0$, for any $\lambda>0$, the marginal distribution of 
$(\widehat{\beta}_{\lambda,j}, \widetilde{\beta}_{\lambda,j})$
is 
\begin{align}
\label{eq:marginal_dist}
\mathcal{N}\left(
\begin{bmatrix}
[\bm{\Sigma}_\lambda^{-1}\bm \Sigma]_{j\cdot}\bm\beta\\
[\bm{\Sigma}_\lambda^{-1}\bm \Sigma]_{j\cdot}\bm\beta 
\end{bmatrix},
\begin{bmatrix}
\sigma^2[\bm{\Sigma}_\lambda^{-1}\bm \Sigma \bm{\Sigma}_\lambda^{-1}]_{jj} & 
\sigma^2([\bm{\Sigma}_\lambda^{-1}\bm \Sigma \bm{\Sigma}_\lambda^{-1}]_{jj} - ([\bm\Sigma_{\lambda}^{-1}]_{jj})^2 s_j)\\
\sigma^2([\bm{\Sigma}_\lambda^{-1}\bm \Sigma \bm{\Sigma}_\lambda^{-1}]_{jj} - ([\bm\Sigma_{\lambda}^{-1}]_{jj})^2 s_j) &
\sigma^2[\bm{\Sigma}_\lambda^{-1}\bm \Sigma \bm{\Sigma}_\lambda^{-1}]_{jj}  
\end{bmatrix}
\right),
\end{align}
which implies that the distribution of 
$(\hat \beta_{\lambda}, \widetilde{\beta}_{\lambda})$ is symmetric, 
as formalized below.
\begin{corollary}
\label{cor:null_W}
Under the same conditions of Proposition~\ref{prop:distribution}, 
when $j$ is a null, the marginal distribution of 
$(\widehat{\beta}_{\lambda,j}, \widetilde{\beta}_{\lambda,j})$  is 
symmetric, i.e., $(\widehat{\beta}_{\lambda,j}, \widetilde{\beta}_{\lambda,j})
\stackrel{\textnormal{d}}{=} (\widetilde{\beta}_{\lambda,j}, \widehat{\beta}_{\lambda,j})$.
\end{corollary}
Several remarks are in order.
\begin{remark}
\label{remark:ols}
For the special case that $\lambda=0$ (OLS), the means and covariances reduce to 
\begin{align}
    &\bm{\mu_{\widehat \beta}} = \bm\beta, \quad 
    \bm{\mu_{\widetilde \beta}} = (\bm I_{p} - \diag\{\bm\Sigma^{-1}\}\bm S)\bm{\beta}, 
    \quad \bm{\Sigma_{\widehat \beta}} = \sigma^2\bm\Sigma^{-1}, \\
    &\bm{\Sigma_{\widetilde \beta}} = \sigma^2 [\bm\Sigma^{-1}- \bm\Sigma^{-1}\bm S \diag\{\bm\Sigma^{-1}\} - \diag\{\bm\Sigma^{-1}\}\bm S \bm\Sigma^{-1} + \\
    &\qquad\qquad\qquad\qquad\qquad \diag\{\bm\Sigma^{-1}\}(\bm{\widetilde X}^\top\bm{\widetilde X}-\bm\Sigma+2\bm S) \diag\{\bm\Sigma^{-1}\}],\\ 
    &\Cov[\bm{\widetilde \beta}_{\lambda=0}, \bm{\widehat \beta}_{\lambda=0}] = \sigma^2(\bm\Sigma^{-1}\bm  - \diag\{\bm\Sigma^{-1}\}\bm S\bm{\Sigma}^{-1}).
\end{align}
In this case, a particularly appealing choice of $\bm S$ is $s_j=1/[\bm \Sigma^{-1}]_{jj}$, which is always allowable (see Section \ref{subsec:generation}). This results in $\bm{\mu_{\widetilde\beta}}=\bm 0_{p\times 1}$ regardless of $\bm\beta$, which should yield better separation between the statistics $\lvert \widehat\beta_{\lambda j} \rvert$ and  $\lvert \widetilde\beta_{\lambda j} \rvert$ when $\beta_j\neq0$. A potentially desirable side effect of this choice for $\bm S$ is that it also results in $ \Cov[\widehat\beta_{\lambda j}, \widetilde\beta_{\lambda j}]=0$
\end{remark}
\begin{remark}
Unlike the BC knockoffs, our knockoffs are not required to satisfy $\bm{\widetilde X}^\top\bm{\widetilde X}=\bm X^\top \bm X$, other than the diagonal conditions $\bm{\widetilde x}_j^\top\bm{\widetilde x}_j=1$. The removal of this condition only affects $\bm{\Sigma_{\widetilde{\beta}}}$ and allows for a more flexible, and potentially better, choice of $\bS$.
As the price, our procedure no longer enjoys provable finite-sample FDR control like the BC procedure. But as we show later, it still delivers asymptotic FDR control 
and finite-sample FDR bounds under mild and verifiable conditions, and it demonstrates good FDR control empirically.

\end{remark}

\subsection{Knockoff generation and choice of $\bm S$}
\label{subsec:generation}

The one-at-a-time knockoff conditions defined in~\eqref{eq:oat_cond} 
are equivalent to requiring
\begin{align}
    &\textnormal{(i)}~
    \bm{X}_{\backslash j}^\top \bm{x}_j = \bm{X}_{\backslash j}^\top \bm{\widetilde x}_j, \quad
    \textnormal{(ii)}~\bm{x}_{j}^\top \bm{\widetilde x}_j = 1 - s_j, \quad 
    \textnormal{{(iii)}}~\bm{\widetilde x}_j^\top \bm{\widetilde x}_j = 1, 
    \quad \textnormal{ for each }j\in[p].\label{eqn:cond}
\end{align}
To provide insight into the role of the knockoffs and how to generate them efficiently, 
consider the singular value decomposition $\bm{X}_{\backslash j}=\bm{U}_{\backslash j}\bm{D}_{\backslash j}\bm{V}_{\backslash j}^\top$, where $\bm{U}_{\backslash j}\in\mathbb{R}^{n\times (p-1)}$ and $\bm{V}_{\backslash j}\in\mathbb{R}^{(p-1)\times(p-1)}$ have orthonormal columns, and $\bm{D}_{\backslash j}\in\mathbb{R}^{(p-1)\times(p-1)}$ is diagonal. Without loss of generality, represent $\bm{\widetilde x}_j$ as
\begin{align}
    \bm{\widetilde x}_j = \bm{U}_{\backslash j}\bm{b}_j + \bm{u}_j b_j + \bm{r}_j \label{eqn:decomposition}
\end{align}
where $\bm{u}_j = (\bm{x}_j - \bm{U}_{\backslash j}\bm{U}_{\backslash j}^\top\bm{x}_j) / \sigma_j$ 
is the unit-norm vector orthogonal to the columns of $\bm{X}_{\backslash j}$ and such that the 
columns of $[\bm{U}_{\backslash j}, \bm{u}_j]$ are an orthonormal basis for the column space of 
$\bm X$, $\sigma_j^2=\lVert \bm{x}_j - \bm{U}_{\backslash j}\bm{U}_{\backslash j}^\top\bm{x}_j \rVert^2 = 1/[\bm{\Sigma}^{-1}_\lambda]_{jj}$ 
is the sum of squares error (SSE) for regressing $\bm{x}_j$ onto $\bm{X}_{\backslash j}$, and 
$\bm{r}_j$ satisfies $\bm X^\top \bm{r}_j=\bm0_{p\times 1}$. 
Our assumption that $\bm X$ is full rank implies $\sigma_j>0$ for $j\in[p]$. To generate $\bm{\widetilde x}_j$, we must find the coefficients $\bm{b}_j$ and $b_j$ and then generate 
$\bm{r}_j$ appropriately. The following proposition characterizes the 
exact form of $\bm{\widetilde{x}}_j$.
\begin{proposition}
\label{prop:kn_form}
For each $j \in [p]$, the $j$-th knockoff satisfying~\eqref{eq:oat_cond} 
must be of the form
\begin{align}
\bm{\widetilde x}_j 
&=\frac{s_j}{\sigma_j^2}\bm{U}_{\backslash j}\bm{U}_{\backslash j}^\top \bm{x}_j + \bigg(1-\frac{s_j}{\sigma_j^2}\bigg) \bm{x}_j + \bm{r}_j, \label{eqn:OATK_knockoff}
\end{align}
where we can select any $s_j \in [0, 2\sigma_j^2]$ 
and then generate $\bm{r}_j$ randomly as any vector orthogonal to the columns of $\bm X$ and with norm $\lVert \bm{r}_j \rVert = (2s_j - s_j^2/\sigma_j^2)^{1/2}$.
\end{proposition}

The proof of Proposition~\ref{prop:kn_form} is relegated to Appendix~\ref{appendix:proof_kn_form}.
Regarding the choice of $\{s_j: j\in[p]\}$, 
Equation~\eqref{eq:marginal_dist} implies that it has no effect on 
the variances of $\{\widetilde\beta_{\lambda j }: j \in [p]\}$, 
although it does affect their correlations (and their correlations 
with $\{\widehat\beta_{\lambda j }: j \in [p]\}$).  
The appealing choice $s_j=\sigma_j^2$ suggested in Remark~\ref{remark:ols} 
is always allowable, and we use this in all of our examples. 
This choice also results in $b_j=0$, in which case \eqref{eqn:OATK_knockoff} becomes
\begin{align}
    \bm{\widetilde x}_j =\bm{U}_{\backslash j}\bm{U}_{\backslash j}^\top \bm{x}_j + \bm{r}_j \label{eqn:OATK_knockoff2}
\end{align}
where $\bm{r}_j$ is  any vector orthogonal to the columns of $\bm X$ 
and with norm $\lVert \bm{r}_j \rVert = \sigma_j$. 
The OATK \eqref{eqn:OATK_knockoff2} is orthogonal to $\bm{u}_j$, the component of $\bm{x}_j$ orthogonal to the column space of $\bm{X}_{\backslash j}$. Consequently, the choice $s_j=\sigma_j^2$ also results in the correlation between $\bm{\widetilde x}_j$ and $\bm{x}_j$ being no more than what ensures that $\bm{\widetilde x}_j$ has the required correlation with $\bm{X}_{\backslash j}$ in \eqref{eqn:cond}.

\subsection{Fast computation of OAT knockoff coefficients}
\label{subsec:fastcomp}
Combining \eqref{eqn:OATK_knockoff} and \eqref{eqn:beta_tilde_vec}, the $j$-th knockoff coefficient for general $s_j$ satisfying $0\leq s_j\leq 2\sigma_j^2$ becomes
\begin{align}
    \widetilde\beta_{\lambda j} &= \widehat\beta_{\lambda j} +[\bm\Sigma_{\lambda}^{-1}]_{jj}\Bigg(\bigg[\frac{s_j}{\sigma_j^2}\bm{U}_{\backslash j}\bm{U}_{\backslash j}^\top \bm{x}_j - \frac{s_j}{\sigma_j^2} \bm{x}_j + \bm{r}_j\bigg]^\top\bm y\Bigg) \\
    &=\widehat\beta_{\lambda j} +[\bm\Sigma_{\lambda}^{-1}]_{jj}\Bigg( -\frac{s_j}{\sigma_j^2} \bigg[\bm{x}_j -\bm{U}_{\backslash j}\bm{U}_{\backslash j}^\top \bm{x}_j\bigg]^\top\bm y + \bm  {r_j}^\top \bm{y} \Bigg) \\
    &= \widehat\beta_{\lambda j} - [\bm\Sigma_{\lambda}^{-1}]_{jj}s_j\widehat\beta_{OLS, j} + [\bm\Sigma_{\lambda}^{-1}]_{jj}\bm{r}_j^\top \bm{y} \label{eqn:knockoff_coefs}
\end{align}
where $\widehat\beta_{OLS, j}$ denotes the coefficient for $\bm{x}_j$ in the OLS regression of $\bm y$ onto $\bm X$. The last equality follows because, by standard partial correlation arguments, $\widehat\beta_{OLS, j}$ is equivalent to the coefficient for the regression of $\bm y$ onto $\bm{x}_j -\bm{U}_{\backslash j}\bm{U}_{\backslash j}^\top \bm{x}_j$, which is
\begin{align}
    \widehat\beta_{OLS, j} = \frac{\big[\bm{x}_j -\bm{U}_{\backslash j}\bm{U}_{\backslash j}^\top \bm{x}_j\big]^\top\bm y}{\big[\bm{x}_j -\bm{U}_{\backslash j}\bm{U}_{\backslash j}^\top \bm{x}_j\big]^\top\big[\bm{x}_j -\bm{U}_{\backslash j}\bm{U}_{\backslash j}^\top \bm{x}_j\big]} = \frac{\big[\bm{x}_j -\bm{U}_{\backslash j}\bm{U}_{\backslash j}^\top \bm{x}_j\big]^\top\bm y}{\sigma_j^2} 
\end{align}
For the specific choice $s_j=\sigma_j^2$, \eqref{eqn:knockoff_coefs} becomes
\begin{align}
    \widetilde\beta_{\lambda j} = \widehat\beta_{\lambda j} -[\bm\Sigma_{\lambda}^{-1}]_{jj}\sigma_j^2\widehat\beta_{OLS, j} +  [\bm\Sigma_{\lambda}^{-1}]_{jj}\bm{r}_j^\top \bm{y} \label{eqn:knockoff_coefs2}
\end{align}
From \eqref{eqn:knockoff_coefs} or the special case \eqref{eqn:knockoff_coefs2}, the knockoff coefficients can be computed efficiently without requiring generation of the full knockoffs or additional regressions that include the knockoffs. One can simply conduct both ridge and OLS regression of $\bm y$ onto $\bm X$ and then compute $[\bm\Sigma_{\lambda}^{-1}]_{jj}\bm{r}_j^\top \bm{y}$ after generating $\bm{r}_j$. An efficient implementation of this procedure and a fast approach for implementing ridge regression for various $\lambda$ and then selecting the best $\lambda$ via leave-one-out cross-validation (LOOCV) is detailed in Appendix \ref{appendix:fast_imp}. In addition to having interpretive value, expression \eqref{eqn:knockoff_coefs2} also enables a very fast version of derandomized knockoffs, which we discuss in Section \ref{subsec:derandom}.

\subsection{Knockoff statistics}
\label{subsec:w}

After constructing $\bm{\widehat \beta}_\lambda$ and $\bm{\widetilde \beta}_\lambda$, the goal is to construct knockoff statistics $ \bm W = [W_1, \dots, W_p]$ such that large positive values of $W_j$ give evidence that $\beta_j\neq0$ for each variable $j\in[p]$. 
Specifically, we focus on 
\begin{equation}
\label{eq:implemented_W}
    W_j = \begin{cases}
        \big|\widehat\beta_{\lambda j}\big| & \text{if } \big|\widehat\beta_{\lambda j}\big| \geq \big|\widetilde\beta_{\lambda j}\big| \\
        -\big|\widetilde\beta_{\lambda j}\big| & \text{if } \big|\widehat\beta_{\lambda j}\big| < \big|\widetilde\beta_{\lambda j}\big|
    \end{cases},
\end{equation}
and our OATK approach selects variables via $\widehat S = \{j: W_j\geq T_\alpha\}$, where the
data-driven threshold $T_\alpha$ is selected (analogous to the BC threshold selection) as
\begin{align}
    \label{eq:W_thresh}
    T_\alpha \equiv\min 
    \Bigg\{ t: \frac{c + \#\{j:W_j\leq -t\}}{1 \vee \#\{j:W_j\geq t\}} \leq \alpha \Bigg\},
\end{align}
where $\alpha \in  (0, 1)$ is the desired FDR and $c$ is the offset parameter, usually fixed as $0$ or $1$. 

The intuition of \eqref{eq:implemented_W} is that a variable is selected only if $\big|\widehat\beta_{\lambda j}\big| \ge T_\alpha$ and $\big|\widehat\beta_{\lambda j}\big| \ge \big|\widetilde\beta_{\lambda j}\big|$. This second inequality reduces false discoveries when there exists strong multicollinearity, which potentially yields large variance of $\widetilde\beta_{\lambda j}$ and therefore large magnitudes of the knockoff coefficient even when the variable is null. We use \eqref{eq:implemented_W} in all our numerical examples due to its empirically accurate FDR control and high power. 

In general, any choice of $\bm W$ can be used with \eqref{eq:W_thresh} 
in our framework as long as they are marginally exchangeable, 
as defined below.

\begin{definition}
    \label{def:W_exchange}
    The statistics $\bm W$ are \textbf{marginally exchangeable} if swapping $\bm{x}_j$ and $\bm{\widetilde{x}}_j$ only switches the sign of $W_j$ for all $j \in [p]$.
\end{definition}

\subsection{Extensions of one-at-a-time knockoffs}
\label{sec:ext}

We introduce three extensions of OATK. The first  
considers multiple exchangeable knockoff copies, 
which allows for producing fine-grained p-values. The second 
adjusts the rejection set of OATK to achieve finite-sample
FDR control with the conditional calibration technique.
The third extension concerns the derandomization of the OATK procedure.

\subsubsection{Multiple OATK}
For each $j\in[p]$, OATK constructs a knockoff copy 
that preserves the correlation structure of the original feature. 
We now consider generating multiple knockoff copies, 
with a joint correlation condition.
Specifically, let $M \ge 1$ denote the number of knockoff copies such that 
$M \le n-p$.
For any $j \in [p]$, let $\bm{\widetilde{x}}_j^{(m)}$ 
denote the $m$-th knockoff copy for $\bm{x}_j$ and
$\widetilde{\bm X}^{(j)} \equiv 
[\bm{\widetilde{x}}_j^{(1)},
\bm{\widetilde{x}}_j^{(2)},
\cdots, \bm{\widetilde{x}}_j^{(M)}]$ denote the assembly of the knockoff copies 
(this is to be distinguished from $\bm{\widetilde X}^{j}$).
We impose the following condition on $\widetilde{\bm X}^{(j)}$:
for each $m \in [M]$,
\begin{align}\label{eq:multiple_cond}
& \text{(i)}~\bm x_i^\top \widetilde{\bm x}_j^{(m)} = \bm x_i^\top \bm x_j, 
~i \in [p]\backslash\{j\}; \text{ (ii)}
~\bm x_j ^\top \widetilde{\bm x}_j^{(m)} = 1 - s_j; \\
& \text{(iii)} ~(\widetilde{\bm x}_j^{(m)})^\top 
\widetilde{\bm x}_j^{(m)} = 1;
~ \text{(iv)} ~(\widetilde{\bm x}_j^{(m)})^\top \widetilde{\bm x}_j^{(\ell)} = 1-s_j,~\ell \in [M]\backslash\{m\},
\end{align}
where $s_j$ is some constant that will be determined shortly. Letting $\bm{e}_M$ be an $M$-dimensional vector of $1$'s,  we generate $\widetilde{\bm X}^{(j)}$ via 
\begin{align}\label{eq:multi_kn} 
\widetilde{\bm X}^{(j)} = \bigg[\frac{s_j}{\sigma_j^2} \bm U_{\backslash j} \bU_{\backslash j}^\top
\bm x_j + \Big(1- \frac{s_j}{\sigma_j^2}\Big) \bm x_j\bigg] \bm e_M^\top
+ \bm R \bm C,
\end{align}
where $\bm R \in \RR^{n \times M}$ is an orthogonal matrix whose column space 
is orthogonal to that of $\bX$ (this is possible since $M \le n-p$), and
$\bm C \in \RR^{M \times M}$ satisfies 
\begin{align}\label{eq:find_c}
\bm C^\top \bm C = s_j \bm I_p + s_j\Big(1 - \frac{s_j}{\sigma^2_j}\Big) \bm e_M \bm e_M^\top.
\end{align}
When $s_j \in [0,\frac{M+1}{M}\sigma^2_j]$, 
the right-hand side of~\eqref{eq:find_c} is positive semi-definite and
such $\bm C$ exists. This is formalized in the following lemma.
\begin{lemma}\label{lem:mult_kn_cons}
Suppose that $n \ge p+M$  and that 
$s_j \le [0,\frac{M+1}{M} \sigma_j^2]$ for each $j\in[p]$. 
Then 
\[
s_j \bm I_p + s_j\Big(1 - \frac{s_j}{\sigma^2_j}\Big)\bm e_M \bm e_M^\top
\]
is positive semi-definite and the knockoff copies defined in~\eqref{eq:multi_kn} satisfy the conditions
in~\eqref{eq:multiple_cond}.
\end{lemma}
The proof of Lemma~\ref{lem:mult_kn_cons} can be found in Appendix~\ref{appx:proof_mult_kn_cons}. 
Compared with~\eqref{eqn:cond}, Condition 
(iv) in~\eqref{eq:multiple_cond} additionally specifies 
the correlation between the knockoff copies, and the conditions are reduced to those of OATK when $M = 1$.

Next, define for each $j\in [p]$ and $m\in[M]$ the design matrix  
${\widetilde{\bm X}}^{j,m} = [\bm{X}_{-j},\bm{\widetilde{x}}_j^{(m)}]$, i.e., 
replacing the $j$-th column by the $m$-th knockoff. 
Let  $\hat \beta_j^{(m)}$ denote the coefficient of $\bm{\widetilde{x}}_j^{(m)}$ 
from regressing $\bm y$ onto $\widetilde{\bm X}^{j,m}$, 
and $\hat \beta_j^{(0)}$ the coefficient of $\bm x_j$ from regressing $\bm y$ onto $\bm X$.
Denoting $M_j = \max \big\{|\hat \beta_j^{(0)}|, |\hat \beta_j^{(1)}|,
\cdots, |\hat \beta_j^{(M)}|\big\}$, we define the OATK p-value as
\begin{align}\label{eq:oatk_pval}
p_j = \frac{1 + \sum_{m=1}^M \mathbf{1}\big\{|\widehat\beta_j^{(m)}| 
\ge |\widehat\beta_j^{(0)}| \big\}}{M+1}.
\end{align}
The selection procedure is based on $\{(M_j,p_j)\}_{j\in[p]}$ and the SeqStep+ filter~\citep{BC_2015}, proceeding in two steps: (1) the features are ordered 
according to $M_j$ such that 
$M_{\pi(1)} \le M_{\pi(2)} \le \cdots \le M_{\pi(p)}$,
where $\pi$ is a permutation of $[p]$;
(2) the rejection set is  $\{\pi(k): k \ge \hat{k}, p_{\pi(k)} \leq \gamma  \}$, where
\begin{align*}
\hat{k} = \inf\left\{k \in[p]: \frac{c+
\# \{j \ge k: p_{\pi(j)} > \gamma\}}{1\vee \#\{j\ge k:p_{\pi(j)} \le \gamma\} } 
\le \frac{1-\gamma}{\gamma} \alpha \right\},
\end{align*}
for some constant $\gamma \in [\frac{\alpha}{\alpha+1},1/2]$.
This selection rule reduces to the basic OATK when $M = 1$.

The following lemma proves that $p_j$ is super-uniform under 
$H_j$. By construction, $p_j$ is supported 
on $\{\frac{1}{M+1},\frac{2}{M+1},\cdots,1\}$, and 
when $M$ is large, the p-value is sufficiently fine-grained to demonstrate the strength of evidence against the null hypothesis.

\begin{lemma}\label{lem:multiple}
Suppose there are no ties among $\{\hat \beta^{(0)}_j, 
\hat \beta^{(1)}_j,\cdots, \hat \beta^{(M)}_j\}$ for 
$j \in [p]$ almost surely.
Then, $p_j$ defined in~\eqref{eq:oatk_pval} is uniform distributed on 
$\{\frac{1}{M+1},\frac{2}{M+1},\ldots,1\}$
for any $j \in \mathcal{H}_0$.
\end{lemma}
The proof of Lemma~\ref{lem:multiple} can be found in 
Appendix~\ref{appx:proof_multiple}. Multiple OATK can be viewed as a generalization of the conditional randomization test p-value \citep{candes2018panning} to the 
fixed design setting, where a (fine-grained) p-value is obtained
by permutation on a synthetically created group.
Like the $M=1$ case, FDR can be approximately controlled when the dependence 
between p-values is relatively small. For simplicity, we focus on proving FDR control in Section~\ref{sec:theory} for the OATK, which is equivalent to the multiple OATK with $M=1$.

\subsubsection{Conditional calibration for exact FDR control}
As alluded to before, our relaxation of the knockoff conditions 
comes at the price of finite-sample FDR control. Although we will 
show in Section \ref{sec:theory}, 
that OATK exhibits approximate finite-sample FDR control, as well as asymptotic FDR control, 
it still may be desirable to achieve exact finite-sample control in some cases, especially 
when type-I errors are costly.

To see why the OATK conditions do not lead to exact FDR control, 
recall that for BC knockoffs, 
\begin{align*}
\fdr = \EE\bigg[\frac{\sum_{j \in \mathcal{H}_0} \ind\{W_j \ge T_\alpha\}}
{\sum_{j \in [p]} \ind \{W_j \ge T_\alpha\}}\bigg]
\le  \alpha \EE \Big[\frac{\sum_{j \in \mathcal{H}_0} \ind\{W_j \ge T_\alpha\}}
{1+\sum_{j \in \mathcal{H}_0} \ind \{W_j \le -T_\alpha\}}\Big] 
\le \alpha,
\end{align*}
where the first inequality holds by  \eqref{eq:W_thresh} with $c=1$, and the last inequality holds if $W_j \,|\, \bm W_{-j} 
\stackrel{\rm d}{=} -W_j \,|\, \bm W_{-j}$, where $\bm W_{\backslash j}$ is $\bm W$ with its $j$-th item removed, for all null features $j$, which is not the case for our method.

A remedy to the FDR violation (both in theory and in practice) is via 
conditional calibration~\citep{fithian2022conditional,luo2022improving}, 
which is efficiently implementable with OATK. 
To conditionally calibrate OATK, 
suppose that for each $j\in[p]$ we have
a statistic $\bm{G}_j$ that is sufficient for the model described by $H_j$ such that if $j$ is null, we can sample 
from $\bm y \,|\, \bm{G}_j$. In this case, 
we could use Monte Carlo to empirically evaluate 
the quantity
\begin{align}
\phi_j(t;\bm{G}_j) \,:=\, 
\EE\bigg[\frac{\ind\{t \cdot  W_j' \ge T_\beta'\}}{c + \sum_{\ell\in[p]}\ind\{W_{\ell}'\le - T_\beta'\}} - 
\frac{\ind\{W_j' \le -T_\beta'\}}{\sum_{\ell \in [p]} \ind\{W_{\ell}' \le -T_\beta'\}} 
\, \bigg| \, \bm{G}_j\bigg], \label{eq:phi_j}
\end{align}
where $\beta \in (0,1)$ is a parameter not necessarily equal to $\alpha$ and $c\in\RR_{\geq 0}$ is a constant.
The recommended choice is to take $\beta\le \alpha$ to yield better power \citep{fithian2022conditional}.

To calibrate OATK, we first run the base OATK procedure to obtain $T_\beta$ and $\bm W$. Then, for each $j$, compute $\hat{t}_j = \sup\big\{t \in \RR_{\ge 0}: \phi_j(t;\bm{G}_j) \le 0 \big\}$ and let  
\begin{equation}\label{eq:eval}
\begin{aligned}
e_j = \frac{p \ind\{\hat{t}_j \cdot W_j\ge T_\beta\}}
{c + \sum_{\ell \in [p]} \ind\{W_\ell \le -T_\beta\}}, 
\quad \text{ if }\phi(\hat t_j; \bm{G}_j) \le 0;\\
e_j = \frac{p \ind\{\hat{t}_j \cdot W_j  > T_\beta\}}
{c + \sum_{\ell \in [p]} \ind\{W_\ell \le -T_\beta\}}, 
\quad \text{ if }\phi(\hat t_j; \bm{G}_j) > 0.
\end{aligned}
\end{equation}

For a fixed $t$, $\phi_j(t; \bm G_j)$ can be computed using Monte Carlo. For each $j$, we sample $\bm y' \sim \bm y \,|\, \bm{G}_j$, run OATK to obtain $W_j', T_\beta'$ (as a function of $\bm X, \bm y'$), evaluate the term inside the expectation in \eqref{eq:phi_j}, and average across the replicates. 
To compute $e_j$, it suffices to evaluate $\phi_j(t; \bm G_j)$ at $t=T_\beta / W_j$ because $\phi_j(T_\beta / W_j ; \bm G_j) \leq 0$ implies $e_j = p / (c + \sum_{\ell \in [p]} \ind\{W_\ell \le -T_\beta\})$ and $\phi_j(T_\beta / W_j ; \bm G_j) > 0$ implies $e_j=0$.

To obtain the rejection set, we apply the eBH procedure \citep{eBH} to the $e_j$'s,
which sorts $e_{\pi(1)} \geq \cdots \geq e_{\pi(p)}$ and computes 
\begin{align*}
\hat{k} = \max\left\{k \in[p]: \frac{ke_{\pi(k)}}{p} \geq \frac{1}{\alpha}\right\}.
\end{align*}
It returns the set $\{\pi(k): k \le \hat{k} \}$. The following theorem proves that $e_j$'s are {\em compound e-values}~\citep{ignatiadis2024compound} and applying 
the eBH procedure to $e_j$'s guarantees FDR control.
Its proof can be found in Appendix~\ref{appx:proof_cc}.
\begin{proposition}\label{prop:cc}
The $e_j$'s defined in~\eqref{eq:eval} satisfies that $\sum_{j \in \mathcal{H}_0} e_j \le p$ 
and applying the eBH procedure to $e_j$'s controls the FDR at the desired level. 
\end{proposition}

\begin{remark}
If we let $\hat t_j = 1$ (instead of as defined as above), $c = 1$, $\beta =\alpha$ when defining the $e_j$'s, then applying the eBH procedure to $e_j$'s is equivalent to 
applying the BC filter to the $W_j$'s~\citep{ren2024derandomised}. 
\end{remark}

The remaining question is the choice of $\bm S_j$. In our setting, 
where $\bm y = \mathcal{N}\bm{(X\beta}, \sigma^2 \bm I)$, a sufficient 
statistics is $\bm{G}_j = (\bm{X}_{-j}^\top \bm{y},\|\bm{y}\|^2)$, and the full distribution of $\bm y \mid  \bm{G}_j$ can be found in~\citet[Section E.2]{luo2022improving}.

\subsubsection{Derandomized one-at-a-time knockoffs}
\label{subsec:derandom}
A drawback of OATK, and of BC and GM, is that the knockoff variable $\bm{\widetilde x}_j$ is stochastic in the sense that $\bm{r}_j$ in Proposition \ref{prop:kn_form} is not unique, so that repeating the algorithm yields different knockoffs and possibly different rejection sets. As for BC, this is alleviated by the derandomization procedure proposed by \cite{derandomized} and \cite{ren2024derandomised}. For OATK, we adopt the derandomization 
scheme in~\citet{derandomized}: we fix a rejection threshold $\eta$ 
and repeat the basic OATK procedure $M$ times, each time using a different random generation of the $\bm{r}_j$ vectors. The rejection set is then the set of variables that were rejected in more than $\eta$ trials. 

\begin{enumerate}
    \item For each replicate $\ell \in [M]$, generate knockoffs $\bm{\widetilde X}^{(\ell)}$ with randomly generated $\bm{r}_j$ vectors (e.g., as Gaussian vectors that are subsequently made orthogonal to $\bm{X}$ and then rescaled to have the required norm) and construct the rejection set $\widehat S^{(\ell)}$ as in the basic OATK procedure.
    \item For each variable $j=1, \hdots, p$, compute the rejection frequency
    \begin{align}
        \Pi_j = \frac{1}{M}\sum_{\ell=1}^M \mathbf{1}\{ j \in \widehat S^{(\ell)} \}.\label{eq:derandom_freq}
    \end{align}
    \item Return the final rejection set as
        $\widehat S = \{j: \Pi_j \geq \eta \} \label{eq:derandom_eta}$.
\end{enumerate}

OATK allows derandomization to be incorporated in a more straightforward and computationally efficient manner that further improves performance. This is in part due to the removal of the BC knockoff requirement that $\bm{\widetilde X}^\top \bm{\widetilde X} =\bm \Sigma-\bm{S}$, which provides improved flexibility and computational efficiency when generating knockoffs, in addition to allowing us to generate knockoffs one-at-a-time. Specifically, each knockoff variable has a deterministic component and a stochastic component, as given by Proposition \ref{prop:kn_form}. For each derandomization replicate, the first two terms in the right-hand side of \eqref{eqn:OATK_knockoff} remain the same, and only the $\bm{r}_j$ term varies. As a result, when computing the knockoff coefficients via \eqref{eqn:knockoff_coefs2}, the term $ \widehat\beta_{\lambda j} -[\bm\Sigma_{\lambda}^{-1}]_{jj}\sigma_j^2\widehat\beta_{OLS, j}$ is computed only once. Then, $M$ copies of $\bm{r}_j$ are generated to compute $[\bm\Sigma_{\lambda}^{-1}]_{jj}\bm{r}_j^\top \bm{y}$, and subsequently $\widetilde \beta_{\lambda, j}$, for all derandomization replicates.

\section{Theoretical guarantees}
\label{sec:theory}
The insight behind the threshold selection of $T_\alpha$ in \eqref{eq:W_thresh} is essentially the same as in the BC procedure. Consider the FDP of OATK:
\begin{align}
    \text{FDP} &= \frac{\#\{j:\beta_j=0 \text{ and } j\in \widehat{S}\}}
     {\#\{j: j\in \widehat{S}\} \vee 1} 
      = \frac{\sum_{j\in \mathcal{H}_0}\ind(W_j \geq T)}
     {\sum_{j\in [p]}\ind(W_j \geq T)}  
     \stackrel{\textnormal{(a)}}{\approx} \frac{\sum_{j\in \mathcal{H}_0}\ind(W_j \leq -T)} 
     {\sum_{j\in [p]}\ind(W_j \geq T)} + o(1) \nonumber \\
     &\leq  \frac{\sum_{j\in [p]}\ind(W_j \leq -T) }
     {\sum_{j\in [p]}\ind(W_j \geq T)} + o(1) \stackrel{\textnormal{(b)}}{\le} \alpha+ o(1)\label{eq:rough_proof} 
\end{align}
where step (a) is approximately true by    
the symmetry of $W_j$ for null variables (Corollary 2.1) under 
mild correlation across features, and step (b) is ensured by the threshold selection \eqref{eq:W_thresh}. The approximation in step (a) carries an $o(1)$ term due to the samples being finite. 
Note that the first inequality in \eqref{eq:rough_proof} should be close to an equality when the non-null variables are sparse, especially if their coefficients are so large that $W_j\geq T$ is much more likely than $W_j \leq -T$ for the non-null variables. 
We more rigorously derive the FDR control guarantees of OATK in the remainder of this section, with asymptotic and non-asymptotic guarantees covered in  
Sections ~\ref{subsec:asymp} and ~\ref{subsec:finite} , respectively.

\subsection{Asymptotic FDR bounds}
\label{subsec:asymp}

The following theorem shows that when the $W_j$'s are reasonably 
weakly dependent on each other asymptotically, the OAT knockoff FDR is controlled as $p_0$, $p_1$, and $n$ all go to infinity in a certain manner. 
\begin{theorem}\label{thm:fdr}
Assume $p\le n$ and $p_0/p \rightarrow \pi_0 \in (0,1)$ as $p_0,p_1 \rightarrow \infty$. 
Define 
\begin{align*}
F_0(t) \equiv \frac{1}{p_0} \sum_{j \in \cH_0}\PP(W_j \ge t),~ 
G_-(t) \equiv \frac{1}{p} \sum_{j \in [p]}\PP(W_j \le -t),~ 
G_+(t) \equiv \frac{1}{p} \sum_{j \in [p]}\PP(W_j \ge t). 
\end{align*}
Assume that $F_0(t)$, $G_-(t)$, and $G_+(t)$ are all continuous in $t$ and that  
there exist constants $c >0$, and $\kappa \in (0,2)$ such that 
\begin{align}\label{eq:cor_cond}
\sum_{j,k \in [p]} \textnormal{Cov}\big(\ind\{W_j \ge t\}, \ind\{W_k \ge t\}\big) 
\le c \cdot p^\kappa, 
\qquad \forall t \in \mathbb{R}.
\end{align}
Then, for any target FDR level $\alpha \in (0,1)$,  
if there exists $t_\alpha \in \RR$ such that $G_-(t_\alpha)/G_+(t_\alpha) < \alpha$, 
then
\begin{align*}
\underset{p_0,p_1, n \rightarrow \infty}{\lim\!\sup}~
\textnormal{FDR} \le \alpha. 
\end{align*}
\end{theorem}
The proof is given in Appendix~\ref{appendix:fdr_proof}. 
The following remark gives an example where the condition in~\eqref{eq:cor_cond} holds.
\begin{remark}
Consider $W_j = |\hat \beta_{j}| - |\widetilde \beta_{j}|$,
for $\hat \beta$ and $\widetilde \beta$ obtained 
from OLS regression (original and knockoff) as described in 
Section~\ref{subsec:structure} with 
$s_j = \sigma_j^2$. Then, we have that 
$\Cov(\hat{\bm{\beta}}, \widetilde{\bm{\beta}}) = 0$, implying that  
$\hat{\bm{\beta}} \,\indep\, \widetilde{\bm{\beta}}$. 
As a result, for any $j,k \in \mathcal{H}_0$, 
$(\hat \beta_j, \hat \beta_k) \,\indep\, (\widetilde \beta_j, \widetilde \beta_k)$.
To proceed, define the maximum correlation between two random variables 
$X$ and $Y$ as 
\begin{align*}
\rho_{\max}(X,Y) = \max_{f_1 \in \mathcal{F}_1, f_2 \in \mathcal{F}_2}
\EE[f_1(X)f_2(Y)],
\end{align*}
where $\mathcal{F}_1 = \{f: \EE[f(X)] = 0, \Var(f(X))=1\}$
and $\mathcal{F}_2 = \{f: \EE[f(Y)] = 0, \Var(f(Y))=1\}$.
Next, for any $t\in\RR$, 
\begin{align*}
\Cov\big(\ind\{W_j \ge t\}, \ind\{W_k \ge t\}\big)
&= \Cov\big(\ind\{|\hat \beta_j| - |\widetilde \beta_j| \ge t\}, 
\ind\{|\hat \beta_k| - |\widetilde \beta_k| \ge t\} \big) \\
&\le \textnormal{Corr}\big(\ind\{|\hat \beta_j| - |\widetilde \beta_j| \ge t\}, 
\ind\{|\hat \beta_k| - |\widetilde \beta_k| \ge t\} \big) \\
& \le\rho_{\max}\big(|\hat \beta_j| -  |\widetilde \beta_j|, 
|\hat \beta_k| - |\widetilde \beta_k|\big),
\end{align*}
where the first inequality follows from the fact that 
$\Var(\ind\{W_j \ge t\}) \le 1$ and $\Var(\ind\{W_k \ge t\})\le 1$; 
the second inequality is due to the definition of $\rho_{\max}$.
By Equation (8) in~\citet{bryc2005maximum}, we have 
\begin{align*}
\rho_{\max}\big(|\hat \beta_j| - |\widetilde \beta_j|,
|\hat \beta_k| -  |\widetilde \beta_k|\big)
& \le \max\big\{\rho_{\max}(\hat \beta_j, \hat \beta_k),
\rho_{\max}(\widetilde \beta_j, \widetilde \beta_k)\big\}.
\end{align*} 
As for the first term above,  
\begin{align*}
\rho_{\max}(\hat \beta_j, \hat \beta_k) 
= \textnormal{Corr}(\hat \beta_j, \hat \beta_k)
= \frac{[\bm \Sigma^{-1}]_{jk}}
{\sqrt{[\bm\Sigma^{-1}]_{jj}[\bm\Sigma^{-1}]_{kk}}}. 
\end{align*}
As for the second term, we have that 
\begin{align*}
\rho_{\max}(\widetilde{\beta}_j, \widetilde{\beta}_k)
= \textnormal{Corr}(\widetilde{\beta}_j, \widetilde{\beta}_k)
= \bigg|\frac{[\bm\Sigma^{-1}]_{jk}}
{\sqrt{[\bm\Sigma^{-1}]_{jj}[\bm\Sigma^{-1}]_{kk}}}
- \sqrt{[\bm\Sigma^{-1}]_{jj} [\bm\Sigma^{-1}]_{kk}}
(\widetilde{\bm x}_j^{\top} \widetilde{\bm x}_k - \bm x_j^{\top}  \bm x_k) \bigg|. 
\end{align*}
Combining the two terms, we can see that Condition~\eqref{eq:cor_cond}
is satisfied when 
\[
\sum_{j,k \in [p]} 
\frac{[\bm\Sigma^{-1}]_{jk}}
{\sqrt{[\bm\Sigma^{-1}]_{jj}[\bm\Sigma^{-1}]_{kk}}}
+ \sqrt{[\bm\Sigma^{-1}]_{jj} [\bm\Sigma^{-1}]_{kk}}
\cdot |\widetilde{\bm x}_j^{\top} \widetilde{\bm x}_k - \bm x_j^{\top} \bm x_k| = O(p^\kappa),
\]
for some $\kappa \in (0,2)$.
\end{remark}

\subsection{Non-asymptotic FDR bounds}
Like the BC knockoffs, the key to the FDR control of OATK is the 
symmetry of the test statistics conditional on the other features.
While the BC knockoff construction enforces such 
conditional symmetry, our relaxed construction does not exactly, so the FDR control of the BC procedure does not directly apply. 
The result in this section characterizes the FDR inflation in terms of the 
violation of conditional symmetry.

To start, we define for any $j \in [p]$ the ``symmetry index''
\begin{align*}
\widehat{\textnormal{SI}}_j = 
\frac{\PP(W_j > 0 \, \big|\, |W_j|, \bm W_{\backslash j})}
{\PP(W_j < 0 \, \big|\, |W_j|, \bm W_{\backslash j})}.
\end{align*}
The following theorem provides an upper bound on the FDR of 
OATK as a function of the $\hat{\si}_j$'s.

\label{subsec:finite}
\begin{theorem}\label{thm:finite_fdr}
If $c = 1$ in the definition of $T_\alpha$ in Equation~\eqref{eq:W_thresh}, then 
for any $\epsilon \ge 0$, the FDR of OATK can be bounded as
\begin{align*}
\fdr \le e^{\epsilon} \cdot \alpha + \PP\Big(\max_{j \in \mathcal{H}_0} \hat{\si}_j > e^{\epsilon}\Big).
\end{align*}
\end{theorem}
The proof of Theorem~\ref{thm:finite_fdr} largely follows the 
leave-one-out technique used in~\citet{barber2020robust} and is 
provided in Appendix~\ref{appendix:proof_finite_fdr}. 
\begin{remark}
If the $W_j$'s satisfy the strict knockoff condition in~\citet{BC_2015},  then
$\hat{\textnormal{SI}_j} \equiv 1$ for $j \in \mathcal{H}_0$, and FDR is 
strictly controlled by taking $\epsilon = 0$ in Theorem~\ref{thm:finite_fdr}.
When the strict knockoff condition is violated only mildly, there exists a small
$\epsilon>0$ such that $\PP(\max_{j\in \mathcal{H}_0} \hat{\textnormal{SI}}_j \ge e^\epsilon)$
is also small, and therefore the FDR is approximately controlled.
\end{remark}

\section{Numerical experiments}
\label{sec:numerical}

In this section we show numerically that, relative to existing FDR-controlling algorithms, OATK exhibits higher power with reasonable finite-sample FDR control. We implement OATK using ridge regression, utilizing the computational speedups described in Section \ref{subsec:fastcomp}. For each simulation example, we fix $n, p, p_1$ and uniformly sample $S$ from $[p]$ such that $|S|=p_1$. We set $p_1=30$ for all experiments. The non-null coefficients are set to $\beta_j = \pm A$, where the sign is sampled uniformly at random and the signal amplitude $A$ is varied across experiments. The remaining entries of $\bm \beta$ are set to zero. The desired false discovery rate is fixed as $\alpha=0.1$. In each simulation example, we assume a model for $\bm X$ and generate a different dataset $(\bm X, \bm y)$ for each replicate by sampling $\bm X \in \mathbb{R}^{n \times p}$ from the model, normalizing each column to have unit norm, and generating $\bm y$ according to the linear model of Equation \eqref{eq:linear_model} with unit variance for the noise.  In all experiments, the regularization parameter $\lambda$ is selected using leave-one-out cross-validation, as described in Appendix \ref{appendix:fast_imp}, and the offset parameter in Equation~\eqref{eq:W_thresh} is set to $c=0$. 
For each example, we conducted $100$ replicates. Section \ref{subsec:gaussian_numerical} assumes $\bm X$ to be multivariate normal with different covariance structures, and Section \ref{subsec:markov_X} considers $\bm X$ sampled from a discrete-time Markov chain. Section \ref{subsec:hiv_numerical} considers real genetic data by fixing $\bm X$ as the set of genetic mutations of the HIV genome (HIV Drug Resistance Database \citep{HIVDatabase}). Using this real $\bm X$ (normalized to have columns with unit norm), we construct a semi-synthetic simulation example by sampling $\bm \beta$ and $\bm y$ as described above, similar to the semi-synthetic genetics study of \cite{GM}. We also conduct a genome-wide association study by taking $\bm y$ to be real experimentally collected susceptibility data of the HIV virus to fosamprenavir, a common treatment drug. Section \ref{subsec:multi_cond_numerical} examines further enhancements of OATK when coupled with multi-bit $p$-value and conditional calibration.

In each simulation example, we include both randomized and derandomized versions of OATK. The existing algorithms with which we compare OATK are the Benjamini-Hochberg procedure \citep{benjamini1995controlling} (BH), BC, and GM. The BC and GM procedures were implemented with Lasso regression, as we observed higher power compared to ridge regression. Additionally, for the Markov chain benchmark in Section \ref{subsec:markov_X}, we compare OATK against the modified knockoff algorithm of \cite{sesia2019} (KnockoffDMC), which generates knockoff variables distributed as a discrete-time Markov chain. Since we only consider low-dimensional datasets where $p<n$, we use the formulation of GM that requires no pre-screening of variables, detailed in Appendix \ref{appendix:sim}. The R package for OATK is publicly available at \url{https://github.com/charlie-guan/oatk}.

\subsection{Simulation with Gaussian $\bm X$}
\label{subsec:gaussian_numerical}

In this example, each row of $\bm X$ is generated independently from a multivariate normal distribution, $\text{N}(0, \bm{\Sigma}_x)$, with three different structures of $\bm{\Sigma}_x$. Namely, we consider
\begin{enumerate}
    \item \textbf{Power decay}: $[\bm \Sigma_x]_{ij} = \rho ^{|i-j|}$ for all $i, j\in[p]$. 
    \item \textbf{Constant positive correlation}: $[\bm \Sigma_x]_{ij} = \rho^{\mathbf{1}(i\neq j)}$ for all $i, j\in[p].$
    \item \textbf{Constant negative correlation}: $\bm \Sigma_x = \bm Q^{-1}$, where $Q_{ij}=\rho^{\mathbf{1}(i\neq j)}$ for all $i, j \in[p]$. 
\end{enumerate}

In all following simulations, we set $\rho=0.4$ for all covariance structures. FDR and power results under different $\rho\in [0, 0.8]$ are included in Appendix \ref{appendix:gaussian}. Changing $\rho$ does not change OATK's uniformly higher power compared to other algorithms.

\begin{figure}[htbp]
    \centering
    \includegraphics[width=0.8\textwidth]{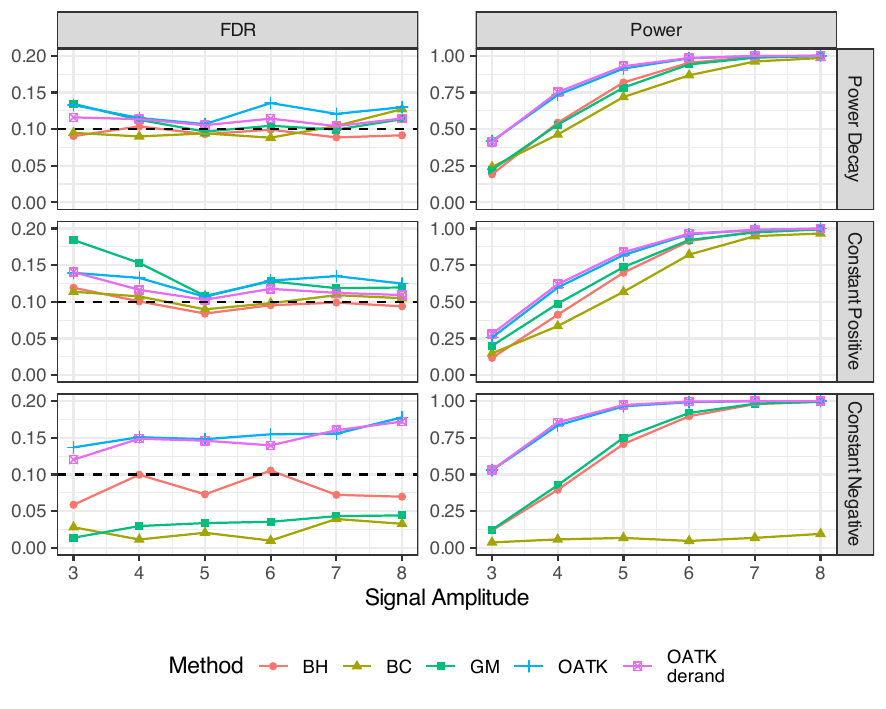}
    \caption{FDR and power as functions of signal amplitude $A$ in the $p=300$ and $n=1000$ Gaussian $\bm X$ study.}
    \label{fig:gaussian-main-small}
\end{figure}

\paragraph{Small $p$ and $n$ study.} We set $n=1000$ and $p=300$. Fig. \ref{fig:gaussian-main-small} plots the FDR and power (the average FDP and TDP, respectively) across the 100 replicates for each example. OATK exhibits significantly higher power than the other algorithms across all three covariance structures and all signal amplitudes. OATK exhibits slight FDR inflation with the power decay and constant positive covariances, but its FDR never exceeded $0.12-0.13$, except in $A=3$. For the constant negative covariance, OATK exhibited somewhat higher FDR inflation at around $0.15$ but far higher power than other algorithms. Derandomization, which is efficiently implementable with OATK due to its substantially lower computational expense, lowers the FDR inflation somewhat with no effect on power. GM typically has somewhat lower power and lower FDR than BH.  In the constant positive covariance at lower amplitudes, GM had higher FDR inflation than all other methods. BC exhibits consistent FDR control and higher power than BH and GM for the power decay and constant positive correlations, but its power drops to almost zero for the constant negative correlation. It should also be noted that when we varied $\rho$, similar conclusions hold regarding the generally superior power and reasonable FDR control of OATK, as we demonstrate in Appendix \ref{appendix:gaussian}.

\begin{figure}[htbp]
    \centering
    \includegraphics[width=0.8\textwidth]{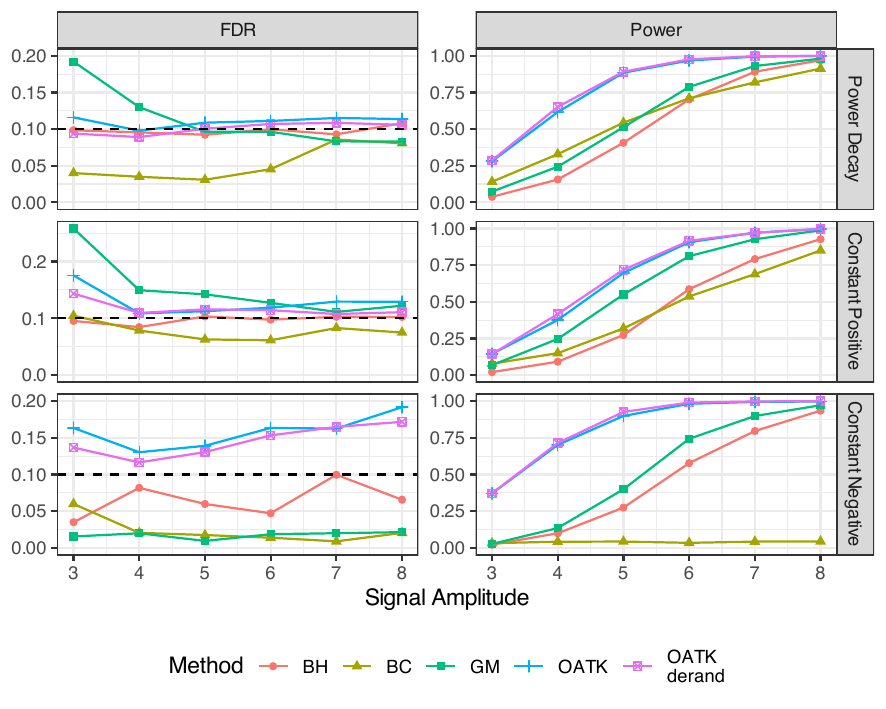}
    \caption{FDR and power as functions of signal amplitude $A$ in the $p=1000$ and $n=2000$ Gaussian $\bm X$ study.}
    \label{fig:gaussian-main-large}
\end{figure}

\paragraph{Larger $p$ and $n$ study.} We next set $n=2000$ and $p=1000$. Fig. \ref{fig:gaussian-main-large} shows OATK again has the best power among all procedures. While it still has relatively mild FDR inflation in some cases, the differences in power between OATK and the other methods are even higher than in the smaller $p$ and $n$ study. Derandomization again lowers the FDR inflation, while slightly boosting power.

\begin{figure}[htbp]
    \centering
    \includegraphics[width=0.8\textwidth]{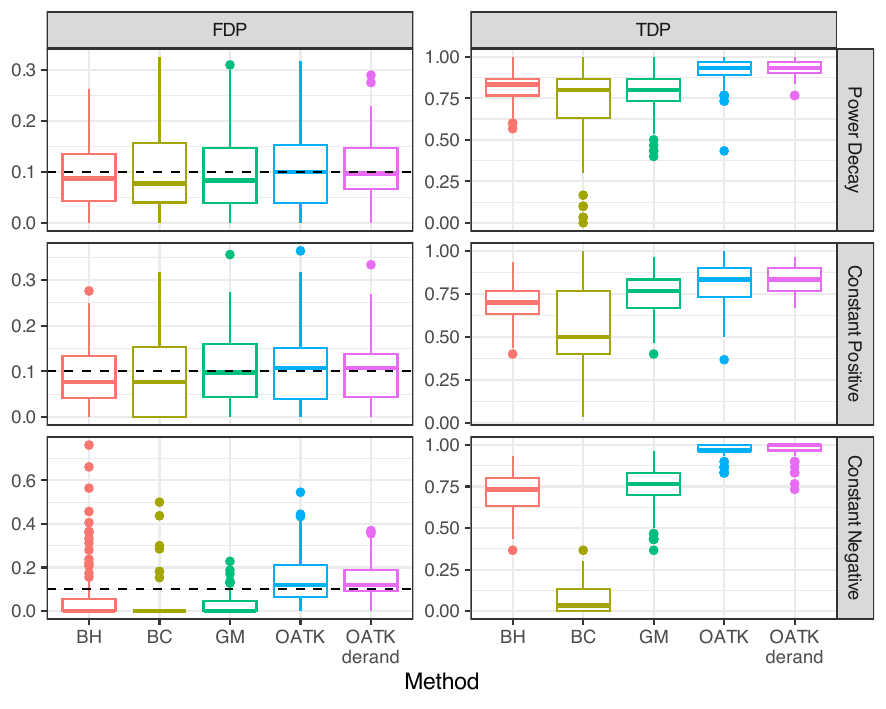}
    \caption{Boxplots of FDP and TDP for individual replicates in the $p=300$ and $n=1000$ Gaussian $\bm X$ study for signal amplitude $A=5$. }
    \label{fig:gaussian-box-small}
\end{figure}

\begin{figure}[htbp]
    \centering
    \includegraphics[width=\textwidth]{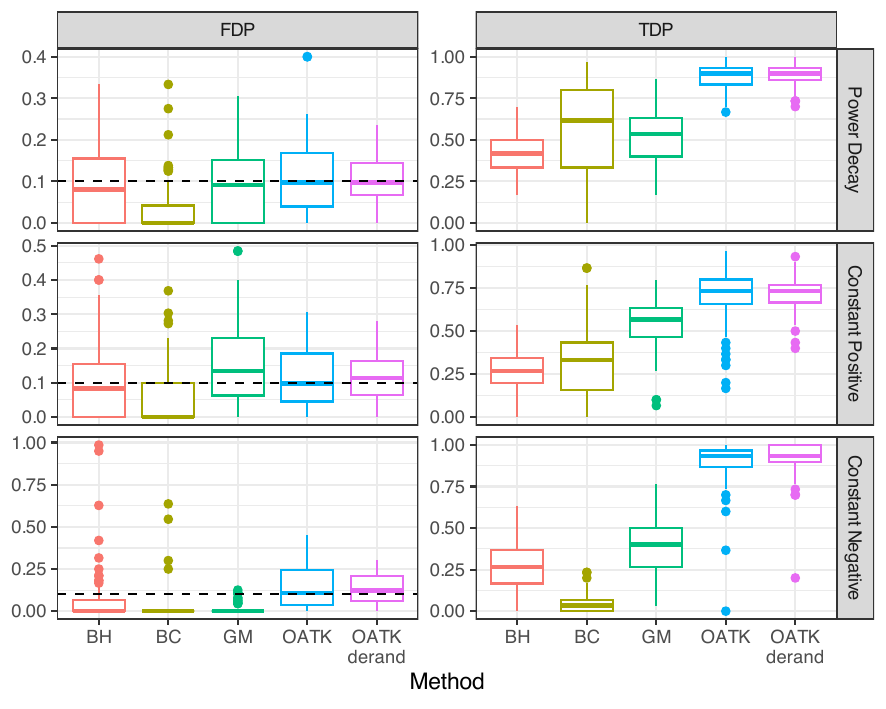}
    \vspace*{-20mm}
    \caption{Boxplots of FDP and TDP for individual replicates in the $p=1000$ and $n=2000$ Gaussian $\bm X$ study for signal amplitude $A=5$.}
    \label{fig:gaussian-box-large}
\end{figure}

Regarding replicate-to-replicate variability, Fig. \ref{fig:gaussian-box-small} and \ref{fig:gaussian-box-large} show the distribution of FDP and TDP across all replicates at a fixed signal amplitude of $A=5$ in the small $p$ and $n$ study and the large $p$ and $n$ study, respectively, from which it can be seen that derandomization lowers the variances of both FDP and TDP in OATK. One drawback of OATK is that it only satisfies asymptotic FDR control, while other methods such as BH and BC satisfy finite-sample FDR control for any $n$. As a result, OATK exhibits some FDR inflation in Figs. \ref{fig:gaussian-main-small} and \ref{fig:gaussian-main-large}, while the two competing procedures do not. GM, which satisfies only asymptotic FDR control, also shows FDR inflation in some cases, especially in the constant positive covariance example. 
However, in light of the replicate-to-replicate variability seen in Figs. \ref{fig:gaussian-box-small} and \ref{fig:gaussian-box-large} for all methods, small inflation of the FDR may be viewed as relatively innocuous, since it is small relative to the FDP variability. In particular, users apply the approach to a single data set, and the actual FDP for that data set will often differ substantially from the FDR due to the FDP variability.  
From Figs. \ref{fig:gaussian-box-small} and \ref{fig:gaussian-box-large}, all methods yield quite high FDPs (e.g., above $0.25$) on at least some replicates, and the upper FDP quartile for even the BC and BH methods (which guarantee $\text{FDR} \le 0.1$) exceeds $0.15$ for many examples. In the constant negative example, BH and BC suffer from more extreme outliers where the FDP is high, and they sometimes yield higher variability of TDP compared to OATK. Moreover, the derandomized OATK reduces the variances of both the FDP and TDP to levels that are lower than the other methods, so that it actually yields less extreme worst-case FDPs and TDPs than the other methods.

\subsection{Simulation with Markov chain design}
\label{subsec:markov_X}

Our OATK approach applies to general fixed-design matrix $\bm X$, as long as it is full rank. Next we consider the setting of \cite{sesia2019}, who developed a knockoff filter specialized to $\bm X$ that is generated from a Markov chain. Specifically, for $i\in[n]$, the $i$-th row $\bm X_{i.}$ of $\bm X$ is generated independently of the other rows as a scalar discrete Markov chain with state space $\mathcal{V} = \{0, 1, 2\}$, as follows. We first sample the Markov chain hyperparameters $\{\gamma_j: j\in[p-1]\} \sim \text{Unif}(0, 0.5)$, with the same hyperparameters applied to each row. We sample the initial state $X_{i,1}$ uniformly at random from $\mathcal{V}$. Then for $j\in[p-1]$, we sample the next state 
$X_{i,j+1}|X_{i,j}$ using the transition probability
\begin{align*}
    Q_j(r \mid s) = \begin{cases} \frac{1}{3} + \frac{2}{3}\gamma_j & r = s \\
    \frac{1}{2}\left(1 - \frac{1}{3} - \frac{2}{3}\gamma_j \right)  & r\neq s
    \end{cases}. 
\end{align*}
In this model, the chain is more likely to remain in the same state than to jump to a different state, but this sojourn probability differs across covariates due to different $\gamma_j$. If the chain does jump to a different state, then it chooses between the other two possible states uniformly at random.  

\begin{figure}[htbp]
    \centering
    \includegraphics[width=0.7\textwidth]{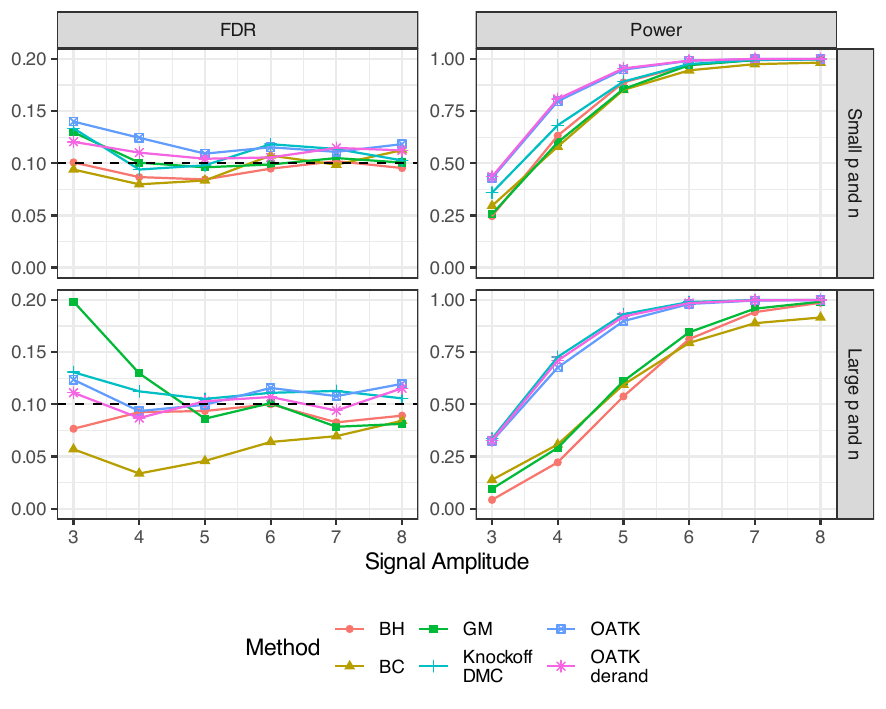}
    \vspace*{-5mm}
    \caption{FDR and power as functions of signal amplitude $A$ in the Markov chain $\bm X$ study for $n=1000$ and $p=300$ (top) and $n=2000$ and $p=1000$ (bottom). }
    \label{fig:markov-chain-line}
\end{figure}

\begin{figure}[htbp]
    \centering
    \includegraphics[width=0.7\textwidth]{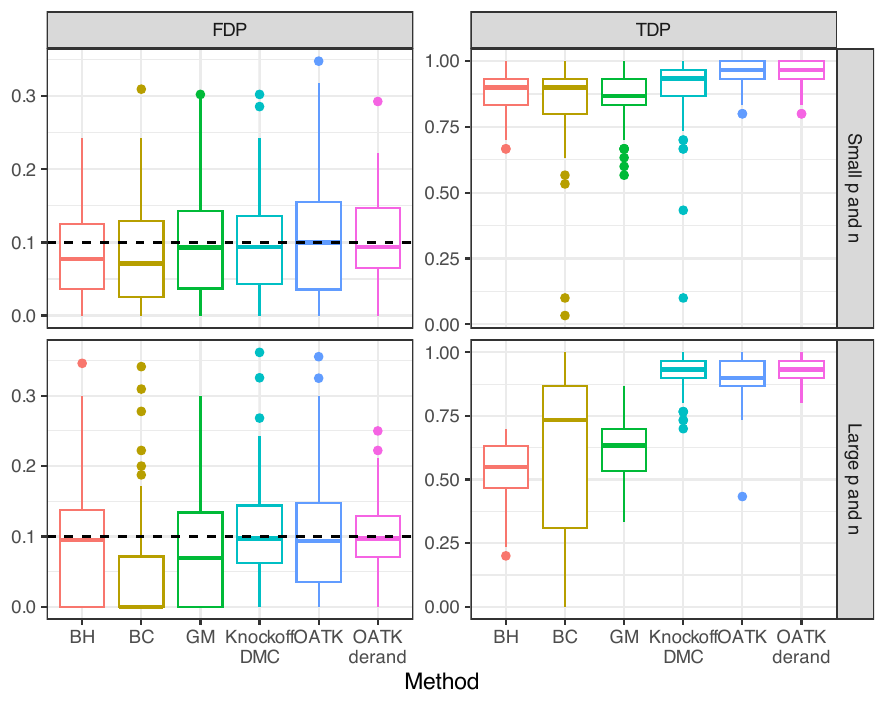}
    \vspace*{-5mm}
    \caption{Boxplots of FDP and TDP for individual replicates in the Markov chain $\bm X$ study with $A=5$ for $n=1000$ and $p=300$ (top) and $n=2000$ and $p=1000$ (bottom).}
    \label{fig:markov-chain-scatter}
\end{figure}

Fig. \ref{fig:markov-chain-line} shows OATK performs the best in power for both smaller $p$ and $n$ ($n=1000, p=300$) and larger $p$ and $n$ ($n=2000, p=1000$), although in the latter situation its power is comparable to KnockoffDMC. In the smaller setting, there is some FDR inflation but derandomization lowers it closer to the target level, and the inflation is comparable to that of other methods, especially KnockoffDMC. In the larger setting, the FDR control is essentially at the target level. Moreover, OATK does not have the significant FDR inflation at lower signal amplitudes that GM does. Relative to KnockoffDMC, OATK performs similarly in power at the larger $p$ and $n$ setting and better in power in the smaller setting, in spite of the fact that KnockoffDMC was developed specifically for, and only applies to, the Markov chain situation and requires estimating the transition probability matrix $Q$ prior to variable selection. Additionally, OATK with derandomization yields the smallest replicate-to-replicate variability in TDP (Fig. \ref{fig:markov-chain-scatter}). KnockoffDMC suffered from poor TDP in several replicates. OATK avoided these extreme low-power outliers, except for one replicate which was avoided via derandomization. Therefore, even when the power is comparable, OATK is a more reliable and consistent variable selection procedure, especially in cases where reproducibility is essential.

\subsection{Genome-wide association study with HIV drug resistance data}

\label{subsec:hiv_numerical}

Genome-wide association studies (GWAS) examine a genome-wide collection of genetic variants from individuals sampled from a target population to detect whether any variants are associated with a phenotype, e.g., assessing which mutations  increase resistance to therapeutics for patients with human immunodeficiency virus (HIV). 

\subsubsection{Semi-synthetic example with real covariates and synthetic response}

\label{subsubsec:hiv_semi}

\begin{figure}[htbp]
    \centering
    \includegraphics[width=0.8\textwidth]{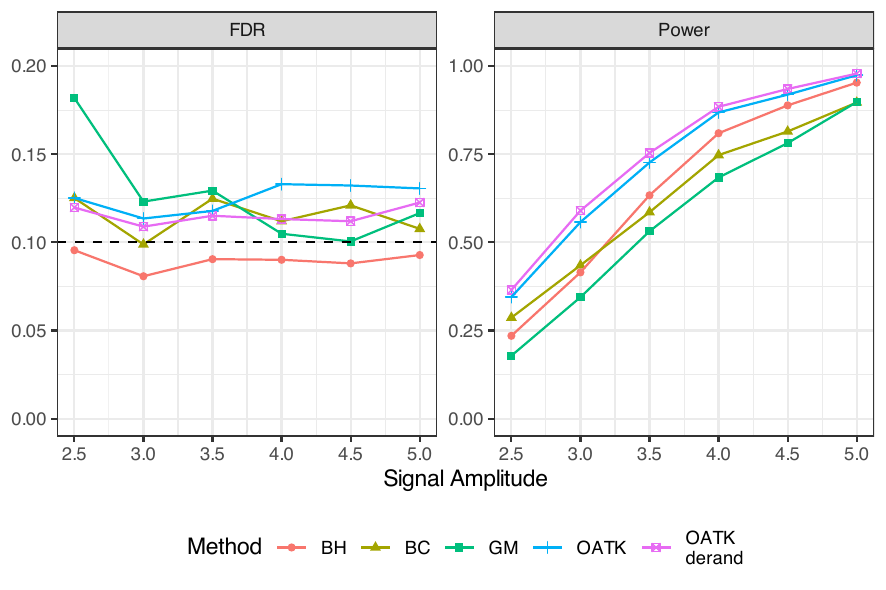}
    \vspace*{-5mm}
    \caption{FDR and power as functions of signal amplitude $A$ for the semi-synthetic HIV drug resistance example.}
    \label{fig:hiv-line}
\end{figure}

We consider real genetic covariates using the HIV Drug Resistance Database of \cite{HIVDatabase}, available at \url{https://hivdb.stanford.edu}. Their protease inhibitor (PI) database collected 2395 genetic isolates. After removing duplicate data, rows with missing values, and rare mutations (i.e., occurring in less than 10 data points), we obtain $\bm X$ with $n=2372$ samples and $p=176$ genetic mutations. The possible values of $X_{ij}$ are 0 or 1, which indicate the absence ($0$) or presence ($1$) of the $j$-th genetic mutation in the $i$-th patient's genome. When constructing $\bm \beta$, a positive non-null coefficient means the presence of its corresponding mutation increases the response phenotype (e.g., increased drug resistance), and a negative coefficient indicates lowered response. The zero coefficients correspond to mutations that have no effect on the response. We set $k=20$. The results are shown in Fig. \ref{fig:hiv-line}. OATK again yields the best power but with some FDR inflation that is reduced by derandomization to levels comparable with BC. All of the methods except BH have some FDR inflation.

\subsubsection{Identification of key genetic mutations that affect resistance to antiretroviral therapy}

In this example, we use completely real data by using the same $\bm X$ as Section \ref{subsubsec:hiv_semi} and taking $\bm y$ as the logarithmic fold resistance to fosamprenavir, a common antiretroviral therapeutic for HIV patients. We do not run any simulations, as both $\bm X$ and $\bm y$ are data collected from real laboratory tests done \textit{in vitro}, as part of the HIV Drug Resistance Database. The experimental methodology used to collect the data is detailed in \cite{HIVDatabase}.
The goal of this study is to compare variable selection procedures for identifying the relevant mutations that impact a patient's drug resistance to fosamprenavir, and corroborate any identified mutations from additional clinical trials.

The data was cleaned as described in Section \ref{subsubsec:hiv_semi}, now resulting in $\bm X$ having $n=2224$ samples and $p=172$ genetic mutations. The dimensionality is slightly different from Section \ref{subsubsec:hiv_semi} due to the addition of the real fold resistance data to fosamprenavir. Table 1 lists the total number of key mutations identified by each procedure, as well as the ones that were uniquely identified (i.e., not identified by any of the other procedures). OATK selected 77 key mutations, which is the most among the four benchmarked algorithms; it also had the highest number of mutations not identified by the other three algorithms. In particular, OATK uniquely identified the mutations 36L, 71V, 73A, and 73C. As an attempt to verify whether these uniquely identified mutations were true or false positives, we found that they were noted by several clinical studies \citep{Lastere, pellegrin, masquelier} to affect virological response in HIV patients when treated with fosamprenavir. We emphasize for the results in Table 1, we used only laboratory tests \textit{in vitro} and did not use clinical data. In this sense, OATK uniquely identified multiple mutations that were corroborated to have clinical impact. In contrast, the uniquely identified mutations of other procedures were not clinically corroborated, to the best of our knowledge.

\begin{table}[htbp]
\label{tab:fpv_genes}
\caption{Variable selection results identifying key genetic mutations that affect HIV drug resistance to fosamprenavir. The number in each mutation label denotes the affected codon position, and the letter is the resulting amino acid from the mutation. Uniquely identified mutations in bold font were verified to be true positives in separate clinical studies.}
\begin{center}
\begin{tabular}{rrrrr}
Procedure & \# Identified & Uniquely Identified Mutations \\\hline
BH & 60 & None  \\
BC & 64 & 12S, 17E, 55R, 66V  \\
GM & 71 & 23I, 37S, 57G, 68E  \\
OATK & 77 & 12P, 13V, \textbf{36L}, 67E, \textbf{71V}, \textbf{73A}, \textbf{73C} \\
\end{tabular}
\end{center}
\end{table}

\subsection{Effect of multi-bit and conditional calibration on OATK}

\label{subsec:multi_cond_numerical}

\begin{figure}[htbp]
    \centering
    \includegraphics[width=0.7\textwidth]{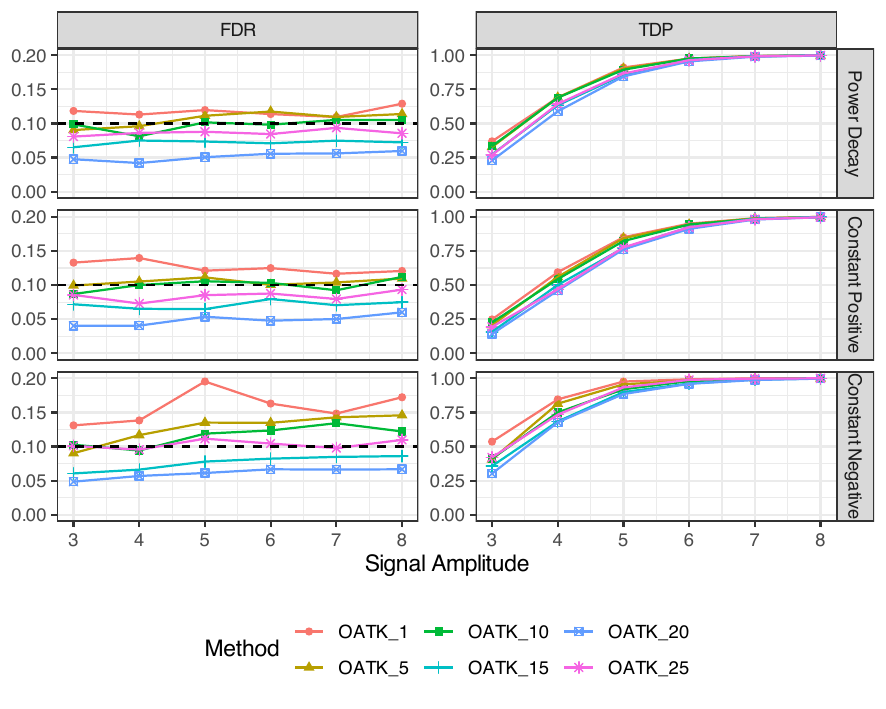}
    \caption{Effect of multi-bit on OATK in the $n=1000$, $p=300$ Gaussian $\bm X$ study. The notation OATK\_M indicates $M$ knockoff variables were simultaneously generated.}
    \label{fig:gaussian-multi-large}
\end{figure}

Fig. \ref{fig:gaussian-multi-large} compares the performance of OATK with ($M>1$) and without ($M=1$) the multi-bit p-value enhancement for various generation sizes $M$ for the same Gaussian $\bm X$ setting of Section \ref{subsec:gaussian_numerical} presented in Figure \ref{fig:gaussian-main-small}. Relative to basic OATK ($M=1$), multi-bit OATK lowers the FDR inflation but reduces power, especially for the constant negative covariance example for which the FDR inflation was higher. However, even with the power reduction, multi-bit OATK still had as high or higher power (and sometimes substantially higher) than the other methods across all three examples (compare Figs. \ref{fig:gaussian-main-small} and \ref{fig:gaussian-multi-large}). For $M=15, 20$, multi-bit OATK conservatively controlled FDR below the specified $\alpha = 0.1$ across all examples and signal amplitudes. Multi-bit OATK with $M=25$ yielded the closest FDR to the desired $\alpha=0.1$ compared to lower $M$'s.

\begin{figure}[htbp]
    \centering
    \includegraphics[width=0.7\textwidth]{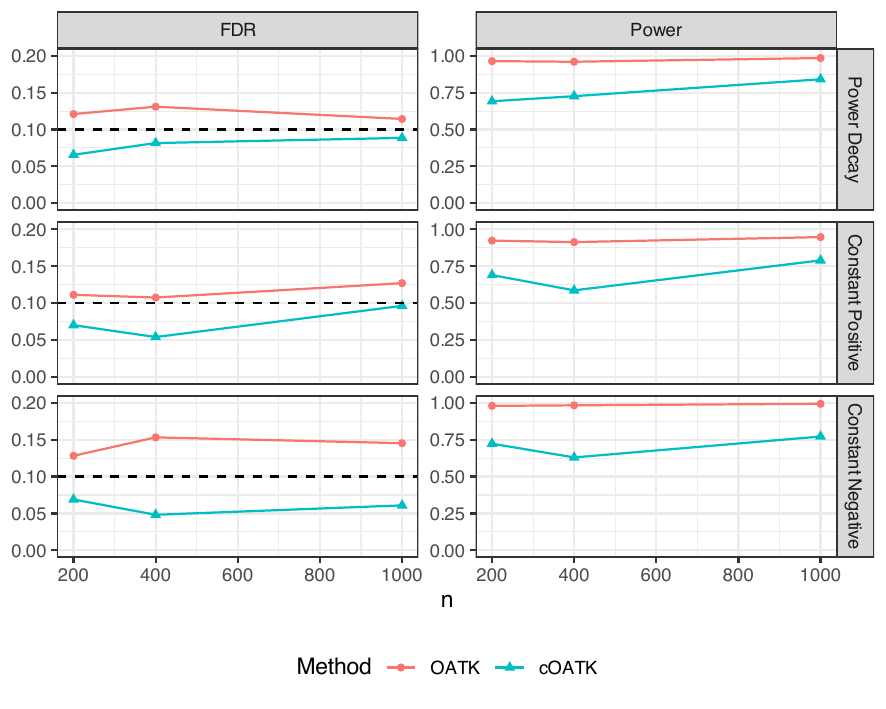}
    \caption{Effect of conditional calibration (cOATK) on OATK for the Gaussian $\bm X$ study with various $n$ with $\rho=0.4, A=6$.}
    \label{fig:calibration}
\end{figure}

Fig. \ref{fig:calibration} shows the effects of conditional calibration on OATK for a Gaussian $\bm X$ example with varying $n$. For  $n=200$, $n=400$, and $n=1000$, we set $p=100$, $p=200$, and $p=300$, respectively. We fix the signal amplitude at $A=6$. Rather than computing $e_j$ for every variable using Monte Carlo simulations, we only carry out the calibration step on promising variables, following a similar approach by \cite{luo2022improving} to improve computational speed. The implementation details are in Appendix \ref{appendix:calibration}. Fig. \ref{fig:calibration} shows the conditionally calibrated OATK corrects for the FDR inflation across all $n$ for all covariance cases without having to generate $e_j$ for all $j$, confirming conditional calibration achieves finite-sample FDR control. However, it also lowers the power compared to the basic OATK.

\section{Conclusions}

This paper introduces one-at-a-time knockoffs, a more powerful and computationally efficient algorithm for controlling the false discovery rate while discovering true signals in OLS and ridge regression problems with $p < n$. It substantially relaxes the conditions knockoffs must satisfy in the original BC knockoff filter, by removing any requirements on the correlation between knockoffs for different variables. This relaxation enables knockoffs to be generated one-at-a-time in a computationally efficient manner and knockoff regressions that involve the exact same number of variables as the original regression. In contrast, the BC knockoff approach requires knockoffs to be generated simultaneously to satisfy the additional requirement $\bm{\widetilde X}^T\bm{\widetilde X}$ and conducts a knockoff regression with twice the number of variables as the original regression, which unnecessarily reduces the power of the knockoff approach.

We also developed a fast version of OATK that avoids having to explicitly generate the knockoff variables $\bm{\widetilde X}$ or compute their coefficients $\bm{\widetilde\beta}$ in full regressions. This allows additional performance-boosting enhancements like derandomization (to reduce variance), multi-bit p-values (to boost power or quantify uncertainty), and conditional calibration (to achieve finite-sample false discovery rate control) to be implemented efficiently. We have proven that OATK controls the FDR at any desired level asymptotically and provided a finite-sample bound on the FDR.
   
One limitation of our current OATK approach is that it is restricted to the case $p<n$ (the original BC approach is restricted even further to $2p<n$). One version of the GM approach, which also tests each variable one at a time, allows $p>n$ via a pre-screening step that uses Lasso regression to select a subset of fewer than $n$ variables, to which the GM procedure is then applied. We implemented similar pre-screening with our OATK approach and observed from the numerical results that it (\textit{i}) works well with $p>n$ and (\textit{ii}) also boosts power in the $p<n$ case while still controlling the FDR. We also found the pre-screening to substantially improve the power of the GM approach when $p<n$. However, its theoretical justification is unclear, and the theoretical results on FDR control in \cite{GM} for the GM approach with $p>n$ do not apply to the actual numerical implementation in their companion code \citep{gm_code} nor to the version of OATK with pre-screening that we observed to work well empirically. We are currently developing an extension of OATK that applies to either $p<n$ or $p>n$ and that has stronger theoretical justification.

\bibliographystyle{apalike}
\bibliography{bibliography}

\newpage
\appendix
\input{appendix}

\end{document}

%% file: appendix.tex
\section{Fast Implementation of One-at-a-time Knockoffs using Ridge Regression}
\label{appendix:fast_imp}

We describe a fast approach for implementing ridge regression for various $\lambda$ and then selecting the best $\lambda$ via leave-one-out cross-validation (LOOCV). Combining LOOCV with Section \ref{subsec:fastcomp}, OATK is far faster than the BC or the GM approach. 

Consider the singular value decomposition $\bm X = \bm{UDV}^\top$, where $\bm U\in\mathbb{R}^n$ and  $\bm V\in\mathbb{R}^p$ have orthonormal columns, and $\bm D=\diag\{d_1, d_2, \hdots, d_p\}$ are the singular values. The OLS and ridge coefficient estimates become
\begin{align}
    \label{eqn:begin_coef}
    &\bm{\widehat\beta}_{OLS} = [\bm X^\top \bm X]^{-1} \bm X^\top \bm y = \bm{VD}^{-2}\bm V^\top\bm{VDU}^\top\bm y = \bm V\diag\bigg\{\frac{1}{d_j}\bigg\}\bm U^\top \bm y\\
    &\bm{\widehat\beta}_{\lambda} =  [\bm X^\top \bm X + \lambda\bm I_{p \times p}]^{-1} \bm X^\top \bm y = [\bm{VD}^{2}\bm V^\top + \lambda\bm{V}\bm V^\top]^{-1}\bm{VDU}^\top\bm y \\
    &\quad =\bm V[\bm{D}^{2} + \lambda]^{-1}\bm V^\top\bm{VDU}^\top\bm y = \bm V\diag\bigg\{\frac{d_j}{d_j^2+\lambda}\bigg\}\bm U^\top \bm y
\end{align}
The vector of fitted response values using $\bm{\widehat\beta}_{\lambda}$ is
\begin{align}
    \bm{\widehat y}_{\lambda} \equiv \begin{bsmallmatrix}
        \widehat y_{\lambda, 1} \\ \vdots \\ \widehat y_{\lambda, n}
    \end{bsmallmatrix} = \bm{X\widehat\beta}_{\lambda} = \bm{UDV}^\top \bm V\diag\bigg\{\frac{d_j}{d_j^2+\lambda}\bigg\}\bm U^\top \bm y = \bm U \diag \bigg\{\frac{d_j^2}{d_j^2+\lambda}\bigg\}\bm U^\top \bm y
\end{align}
and the corresponding LOOCV error sum of squares is
\begin{align}
    \label{eqn:end_coef}
    \SSE_{\lambda, CV} = \sum_{i=1}^n \Bigg(\frac{y_i-\widehat y_{\lambda i}}{1-\big[ \bm U\diag\{\frac{d_j^2}{d_j^2+\lambda}\}\bm U^\top \big]_{ii}}\Bigg)^2 = \sum_{i=1}^n \Bigg(\frac{y_i-\widehat y_{\lambda i}}{1-\bm{U}_{i\cdot}\diag\{\frac{d_j^2}{d_j^2+\lambda}\}\bm{U}_{i\cdot}^\top }\Bigg)^2.
\end{align}
 Equations \eqref{eqn:begin_coef}-\eqref{eqn:end_coef} provide a fast method to fit ridge regression models with many different $\lambda$ values and select the one that minimizes $\SSE_{\lambda, CV}$.

To efficiently compute the knockoff coefficients $\bm{\widetilde\beta}_\lambda$, we use \eqref{eqn:begin_coef} with \begin{align}
    \bm\Sigma_\lambda^{-1} = \bm V \diag\bigg\{\frac{1}{d_i^2+\lambda}\bigg\}\bm V^\top,
\end{align}
so that 
\begin{align}
    [\bm\Sigma_{\lambda}^{-1}]_{jj} &=  \bm V_{j\cdot} \diag\bigg\{\frac{1}{d_i^2+\lambda}\bigg\}\bm V_{j\cdot}^\top \\
    \sigma_j^2 &= \frac{1}{[\bm\Sigma^{-1}]_{jj}} = \frac{1}{\bm V_{j\cdot} \diag\bigg\{\frac{1}{d_i^2}\bigg\}\bm V_{j\cdot}^\top}
\end{align}
To generate the $\bm{r}_j$ vectors required in \eqref{eqn:knockoff_coefs2}, we first generate $\bm{r}_j$ as any random vector (e.g., $\sim\mathcal{N}_n(\bm 0_{n\times 1}, \bm I_{n\times n})$, then update $\bm{r}_j-\bm{UU}^\top\bm{r}_j$, then rescale each $\bm{r}_j$ to have unit norm. Using this unit-norm version of $\bm{r}_j$ in \eqref{eqn:knockoff_coefs2} gives
\begin{align}
    \widetilde\beta_{\lambda j} = \widehat\beta_{\lambda j} -[\bm\Sigma_{\lambda}^{-1}]_{jj}\sigma_j^2\widehat\beta_{OLS, j} + [\bm\Sigma_{\lambda}^{-1}]_{jj}\sigma_j\bm{r}_j^\top \bm{y} .
    \label{eqn:beta_imp}
\end{align}

\section{Technical Proofs}

\subsection{Proof of Proposition~\ref{prop:distribution}}
\label{appendix:proof_distribution}
By standard results for ridge regression, we have 
$\bm{\widehat\beta}_\lambda = \bm{\Sigma}_\lambda^{-1}\bX^\top \bm{y}$ and 
\begin{align}
    &\bm{\mu_{\widehat\beta}} \equiv \mathbb{E}[\bm{\widehat\beta}] = \bm{\Sigma}_\lambda^{-1}\bm{\Sigma\beta} \label{eqn:begin_joint} \\
    &\bm{\Sigma_{\widehat\beta}} \equiv \Cov[\bm{\widehat\beta}] = \sigma^2\bm{\Sigma}_\lambda^{-1}\bm X^\top \bm X \bm{\Sigma}_\lambda^{-1} = \sigma^2\bm{\Sigma}_\lambda^{-1}\bm \Sigma \bm{\Sigma}_\lambda^{-1}.
\end{align}
For the OATK coefficients, 
\begin{align}
    \bm{\widetilde\beta}_\lambda &= \bm{\widehat\beta}_\lambda + \diag\{\bm\Sigma_\lambda^{-1}\}(\bm{\widetilde X}^\top - \bm X^\top)\bm y = \bm\Sigma_\lambda^{-1}\bm X^\top \bm y + \diag\{\bm\Sigma_\lambda^{-1}\}(\bm{\widetilde X}^\top - \bm X^\top)\bm y \\
    &= (\bm\Sigma_\lambda^{-1}\bm X^\top + \diag\{\bm\Sigma_\lambda^{-1}\}(\bm{\widetilde X}^\top - \bm X^\top))(\bm{X\beta}+\bm z) \sim \mathcal{N}_p(\bm{\mu_{\widetilde \beta}}, \bm{\Sigma_{\widetilde \beta}}), 
\end{align}
where 
\begin{align}
    \bm{\mu_{\widetilde \beta}}  &\equiv \mathbb{E}[\bm{\widetilde\beta}] = (\bm\Sigma_\lambda^{-1}\bm X^\top + \diag\{\bm\Sigma_\lambda^{-1}\}(\bm{\widetilde X}^\top - \bm X^\top))\bm{X\beta}\\
    &= (\bm\Sigma_\lambda^{-1}\bm X^\top \bm X + \diag\{\bm\Sigma_\lambda^{-1}\}(\bm{\widetilde X}^\top \bm X - \bm X^\top \bm X))\bm{\beta} =  (\bm\Sigma_\lambda^{-1}\bm\Sigma - \diag\{\bm\Sigma_\lambda^{-1}\}\bm S)\bm{\beta} \label{eqn:mean_tilde} \\
    \bm{\Sigma_{\widetilde \beta}} &\equiv \Cov[\bm{\widetilde\beta}] \\
    &= \sigma^2 (\bm\Sigma_\lambda^{-1}\bm X^\top + \diag\{\bm\Sigma_\lambda^{-1}\}(\bm{\widetilde X}^\top - \bm X^\top)) (\bm\Sigma_\lambda^{-1}\bm X^\top + \diag\{\bm\Sigma_\lambda^{-1}\}(\bm{\widetilde X}^\top - \bm X^\top))^\top \\
    &= \sigma^2 [\bm\Sigma_\lambda^{-1}\bm \Sigma \bm\Sigma_\lambda^{-1} + \bm\Sigma_\lambda^{-1}\bm X^\top (\bm{\widetilde X} - \bm X^) \diag\{\bm\Sigma_\lambda^{-1}\} +\\
    &\quad \diag\{\bm\Sigma_\lambda^{-1}\}(\bm{\widetilde X}^\top - \bm X^\top)\bm X \bm\Sigma_\lambda^{-1} + \diag\{\bm\Sigma_\lambda^{-1}\}(\bm{\widetilde X}^\top - \bm X^\top)(\bm{\widetilde X} - \bm X)\diag\{\bm\Sigma_\lambda^{-1}\}] \\
    &= \sigma^2 [\bm\Sigma_\lambda^{-1}\bm \Sigma \bm\Sigma_\lambda^{-1} - \bm\Sigma_\lambda^{-1}\bm S \diag\{\bm\Sigma_\lambda^{-1}\} - \diag\{\bm\Sigma_\lambda^{-1}\}\bm S \bm\Sigma_\lambda^{-1} + \\
    &\quad\quad\quad \diag\{\bm\Sigma_\lambda^{-1}\}(\bm{\widetilde X}^\top\bm{\widetilde X}-\bm\Sigma+2\bm S) \diag\{\bm\Sigma_\lambda^{-1}\}]
\end{align}
so that the marginal variances of $\widehat\beta_{\lambda j}$ and $\widetilde\beta_{\lambda j}$ are given by
\begin{align}
    \label{eqn:marginal_var}
    \diag\{\bm{\Sigma_{\widehat \beta}}\} = \diag\{\bm{\Sigma_{\widetilde \beta}}\} = \sigma^2 \diag\{\bm\Sigma_\lambda^{-1}\bm \Sigma \bm\Sigma_\lambda^{-1}\}.
\end{align}
We also have
\begin{align}
    \Cov[\bm{\widetilde \beta}_\lambda, \bm{\widehat \beta}_\lambda] &= \mathbb{E}\big[ (\bm\Sigma_\lambda^{-1}\bm X^\top + \diag\{\bm\Sigma_\lambda^{-1}\}(\bm{\widetilde X}^\top - \bm X^\top))\bm z \bm z^\top(\bm X\bm{\Sigma}_\lambda^{-1})\big] \\
    &= \sigma^2(\bm\Sigma_\lambda^{-1}\bm \Sigma \bm\Sigma_\lambda^{-1} - \diag\{\bm\Sigma_\lambda^{-1}\}\bm S\bm{\Sigma}_\lambda^{-1}) ) \label{eqn:end_joint}
\end{align}

\subsection{Proof of Proposition~\ref{prop:kn_form}}
\label{appendix:proof_kn_form}
Recall that we can decompose the OATK as 
\begin{align}\label{eq:representation_again}
    \bm{\widetilde x}_j = \bm{U}_{\backslash j}\bm{b}_j + \bm{u}_j b_j + \bm{r}_j.
\end{align}
With the above representation, Condition (i) in~\eqref{eqn:cond} becomes
\begin{align}
     \bm{X}_{\backslash j}^\top \bm{x}_j = \bm{X}_{\backslash j}^\top \bm{\widetilde x}_j = \bm{V}_{\backslash j}\bm{D}_{\backslash j}\bm{U}_{\backslash j}^\top(\bm{U}_{\backslash j}\bm{b}_j + \bm{u}_j b_j + \bm z_j) = \bm{V}_{\backslash j}\bm{D}_{\backslash j}\bm{b}_j
\end{align}
so that
\begin{align}
     \bm{b}_j = \bm{D}_{\backslash j}^{-1}\bm{V}_{\backslash j}^\top\bm{X}_{\backslash j}^\top \bm{x}_j =\bm{D}_{\backslash j}^{-1}\bm{V}_{\backslash j}^\top\bm{V}_{\backslash j}\bm{D}_{\backslash j}\bm{U}_{\backslash j}^\top \bm{x}_j = \bm{U}_{\backslash j}^\top \bm{x}_j. \label{eqn:b_j}
\end{align}

Notice that $\sigma_j=0$ if and only if $\bm{x}_j$ lies in the column space of $\bm{X}_{\backslash j}$, in which case the term $\bm{u}_j b_j$ in \eqref{eq:representation_again} disappears, 
and we can represent $\bm{\widetilde x}_j = \bm{U}_{\backslash j}\bm{b}_j + \bm{r}_j$,
where $\bm{U}_{\backslash j}\bm{b}_j = \bm{U}_{\backslash j}\bm{U}_{\backslash j}^\top \bm{x}_j= \bm{x}_j$ by~\eqref{eqn:b_j}. 
In this case, Condition (iii) in~\eqref{eqn:cond} implies 
$\lVert \bm{r}_j\rVert=0$, and the only knockoff that satisfies Condition (i) 
and Condition (iii) is $\bm{\widetilde x}_j=\bm{x}_j$. 
This is the primary reason we must assume $\bm X$ is full rank. 

Next, substituting~\eqref{eqn:b_j} into~\eqref{eq:representation_again}, 
Condition (iii) in~\eqref{eqn:cond} becomes 
\begin{align}
    1 &= \bm{\widetilde x}_j^\top \bm{\widetilde x}_j 
    = [\bm{U}_{\backslash j}\bm{U}_{\backslash j}^\top \bm{x}_j + \bm{u}_j b_j + \bm{r}_j]^\top[ \bm{U}_{\backslash j}\bm{U}_{\backslash j}^\top \bm{x}_j + \bm{u}_j b_j + \bm{r}_j] \\
    &= \lVert \bm{U}_{\backslash j}\bm{U}_{\backslash j}^\top \bm{x}_j \rVert^2 + \lVert \bm{u}_j b_j \rVert^2 + \lVert \bm{r}_j \rVert^2 \\
    &= (1-\sigma^2_j)+ b_j^2 + \lVert \bm{r}_j \rVert^2
\end{align}
or $\lVert \bm{r}_j \rVert^2 = \sigma^2_j - b_j^2$, 
which also implies that we need
$\lvert b_j\rvert \leq \lvert\sigma_j\rvert$.
Using the preceding, Condition (ii) in Equation~\eqref{eqn:cond} 
becomes
\begin{align}
    1-s_j &= \bm{x}_{j}^\top \bm{\widetilde x}_j 
    = \bm{x}_{j}^\top[\bm{U}_{\backslash j}\bm{U}_{\backslash j}^\top \bm{x}_j 
    + \bm{u}_j b_j + \bm{r}_j] \\
    &= \lVert \bm{U}_{\backslash j}\bm{U}_{\backslash j}^\top \bm{x}_j \rVert^2 + \frac{\bm{x}_{j}^\top(\bm{x}_j - \bm{U}_{\backslash j}\bm{U}_{\backslash j}^\top\bm{x}_j)b_j}{\sigma_j} \\
    &= (1-\sigma^2_j) + \sigma_j b_j,
\end{align}
or $b_j=\sigma_j-s_j/\sigma_j$, 
which requires
$0\leq s_j\leq2\sigma_j^2$.
Finally, 
\begin{align}
    \lVert \bm{r}_j \rVert^2 = \sigma_j^2-\left(\sigma_j-\frac{s_j}{\sigma_j}\right)^2 = 2s_j - \frac{s_j^2}{\sigma_j^2}.
\end{align}

\subsection{Proof Lemma~\ref{lem:mult_kn_cons}}
\label{appx:proof_mult_kn_cons}
Define $\bm{A}_j = s_j \bm I_p + s_j(1-{s_j}/{\sigma^2_j})  \bm e_M  \bm e_M^\top$.
It can be checked that $\bm A_j$ has two eigenvalues: $s_j(1-s_j/\sigma_j^2)M+s_j$ 
and $s_j$, where the former corresponds to a unique eigenvector and the latter has multiplicity $M-1$. It is then straightforward that $\bm A_j$ 
is positive semi-definite when $s_j \in [0,\frac{M+1}{M}\sigma_j^2]$.

Next, we verify that the knockoff copies constructed in~\eqref{eq:multi_kn}
satisfy the conditions in~\eqref{eq:multiple_cond}. 
For condition (i), we have 
\begin{align}
\bX_{\backslash j}^\top \widetilde{\bX}^{(j)}
= \bigg[\frac{s_j}{\sigma_j^2}\bX_{\backslash j}^\top \bU_{\backslash j}
\bU_{\backslash j}^\top \bm x_j + \Big(1-\frac{s_j}{\sigma_j^2}\Big)\bX_{\backslash j}^\top \bm x_j \bigg]  \bm e_M^\top = \bX_{\backslash j}^\top \bm x_j  \bm e_M^\top.
\end{align}
For condition (ii), there is 
\begin{align}
\bm x_j^\top \widetilde{\bX}^{(j)}
= 
\bigg[\frac{s_j}{\sigma_j^2} (1-\sigma_j^2) + \Big(1-\frac{s_j}{\sigma_j^2}\Big)\bigg] \bm e_M^\top = (1-s_j) \bm e_M^\top.
\end{align}
Finally, we check 
\begin{align}
(\widetilde{\bX}^{(j)})^\top \widetilde{\bX}^{(j)}
& =  \bm e_M \bigg[\frac{s_j}{\sigma_j^2}\bU_{\backslash j}
\bU_{\backslash j}^\top \bm x_j + \Big(1-\frac{s_j}{\sigma_j^2}\Big) \bm x_j \bigg]^\top \bigg[\frac{s_j}{\sigma_j^2} \bU_{\backslash j}
\bU_{\backslash j}^\top \bm x_j + \Big(1-\frac{s_j}{\sigma_j^2}\Big) \bm x_j \bigg] \bm e_M^\top + \bm C^\top \bm C\\
& = \bigg[1-\sigma^2_j + \Big(1-\frac{s_j}{\sigma_j^2}\Big)^2 \sigma_j^2\bigg]  \bm e_M \bm e_M^\top  
+ s_j \bm I_p + s_j\Big(1-\frac{s_j}{\sigma_j^2}\Big)  \bm e_M  \bm e_M^\top \\
& = s_j \bm I_p + (1-s_j) \bm e_M  \bm e_M^\top,
\end{align}
which verifies conditions (iii) and (iv). We have therefore completed the 
proof.

\subsection{Proof of Lemma~\ref{lem:multiple}}
\label{appx:proof_multiple}
Given a permutation $\pi$ on $\{0,1,\cdots,M\}$ and any $j\in \mathcal{H}_0$, 
we first show that 
\begin{align}\label{eq:perm_inv}
(\widehat \beta^{(0)}_j, \widehat \beta^{(1)}_j,\ldots, \widehat \beta^{(M)}_j)
\stackrel{\textnormal{d}}{=}(\widehat \beta^{(\pi(0))}_j,
\widehat \beta^{(\pi(1))}_j,\ldots, \widehat \beta^{(\pi(M))}_j).
\end{align}
Recall that $\widehat \beta^{(m)}_j = f\big((\widetilde{\bm X}^{j,m})^\top
\widetilde{\bm X}^{j,m}, (\widetilde{\bm X}^{(j,m)})^\top \bm y \big)$ for some function $f$.  
By construction, $(\widetilde{\bm X}^{j,m})^\top
\widetilde{\bm X}^{j,m} = \bm X^\top \bm X$ and 
$\bm y^\top \widetilde{\bm X}^{(j,m)} = (\bm y^\top \bm X_{\backslash j}, 
\bm y^\top \widetilde{\bm x}_j^{(m)})$, where we write 
$\widetilde{\bm x}_j^{(0)} = \bm x_j$.
It then suffices to prove that 
\begin{align*}
(\bm y^\top \bm{X}_{\backslash j}, ~\bm y^\top \bm{x}_j,
\bm y^\top \bm{\widetilde{x}}_j^{(1)},  
\ldots,~\bm y^\top \bm{\widetilde{x}}_j^{(M)})
\stackrel{\textnormal{d}}{=} 
(\bm y^\top \bm{X}_{\backslash j}, ~\bm y^\top \bm{\widetilde{x}}_j^{(\pi(0))},
\bm y^\top \bm{\widetilde{x}}_j^{(\pi(1))},  
\ldots,~\bm y^\top \bm{\widetilde{x}}_j^{(\pi(M))}).
\end{align*}
For any $m \in [M]$, we can explicitly write
\begin{align*}
\bm y^\top \bm{\widetilde{x}}_j^{(m)}
= & \bigg(\sum_{k\neq j}\bm{x}_k\beta_k\bigg)^\top \bm{\widetilde{x}}_j^{(m)} + 
\bm z^\top \bm{\widetilde{x}}_j^{(m)} 
= \sum_{k\neq j}\beta_k \bm x_k^\top \bm{\widetilde{x}}_j^{(m)} + 
\bm z^\top \bm{\widetilde{x}}_j^{(m)} \\ 
= & \sum_{k\neq j} \beta_k \bm{x}_k^\top \bm{x}_j + 
\bm z^\top \bm{\widetilde{x}}_j^{(m)} 
= \bm \beta^\top \bm X^\top \bm{x}_j+ \bm z^\top \bm{\widetilde{x}}_j^{(m)},
\end{align*}
where the second-to-last step is due to the construction of $\bm{\widetilde{x}}_j^{(m)}$,
and the last step is because $\beta_j = 0$ under the null.
We then have 
\begin{align}
(\bm y^\top \bm{X}_{\backslash j}, ~ \bm y^\top \bm{x}_j^,~ 
\bm y^\top \bm{\widetilde{x}}_j^{(1)}, 
\cdots,\bm y^\top \bm{\widetilde{x}}_j^{(m)})^\top \sim \mathcal{N}(\bm \mu^{(j)}, \bm \Sigma^{(j)}),
\end{align}
where $(\bm \mu^{(j)})^\top = 
(\bm\beta^\top \bm X^\top \bm X_{\backslash j}, 
\bm\beta^\top \bm X^\top \bm x_{j},
\cdots, \bm\beta^\top \bm X^\top \bm x_j)$ and 
\begin{align}
\bm \Sigma^{(j)} = 
\begin{pmatrix}
\bm \Sigma_{\backslash j, \backslash j} & \bm \Sigma_{\backslash j, j} \bm e_{M+1}^\top \\
\bm e_{M+1} \bm \Sigma_{j, \backslash j} & (1-s_j)\bm e_{M+1} \bm e_{M+1}^\top + 
s_j \bm I_{M+1}
\end{pmatrix},
\end{align}
where $\bm e_{M+1}$ is an $(M+1)$-dimensional vector with 
all entries being one. It can be seen that the above distribution 
is invariant to the
permutation of $\bm{\bm{x}}_j,\bm{\widetilde{x}}_j^{(1)},\ldots,\bm{\widetilde{x}}^{(M)}_j$.
We have thus shown~\eqref{eq:perm_inv}. 

Then for any $m \in \{0,1,\ldots,M\}$, 
\begin{align}
\PP\Big(p_j = \frac{m}{M+1}\Big) 
& = \PP\Big(\text{rank}\big(|\hat \beta_j^{(0)}|\big) = m\Big)\\
& = \frac{1}{(M+1)!}\sum_{\pi: \text{permutation on }[M+1]}
\PP \Big(\text{rank}\big(|\hat \beta_j^{(\pi(0))}|\big) = m\Big)\\
& = \EE\Bigg[\frac{1}{(M+1)!}
\sum_{\pi: \text{permutation on }[M+1]}
\ind\Big\{\text{rank}\big(|\hat \beta_j^{(\pi(0))}|\big) = m\Big\}
\Bigg]\\
& = \frac{M! }{(M+1)!} = \frac{1}{M+1}.
\end{align}
We have therefore completed the proof.

\subsection{Proof of Proposition~\ref{prop:cc}}
\label{appx:proof_cc}
If $\phi_j(\hat t_j; \bm{S}_j) \le 0$, then 
\begin{align*} 
\EE[e_j] & = \EE\bigg[\frac{p \ind\{\hat{t}_j \cdot W_j\ge T_\beta\}}
{c + \sum_{\ell \in [p]} \ind\{W_\ell \le -T_\beta\}}\bigg] \\
& = \EE\bigg[\frac{p \ind\{\hat{t}_j \cdot W_j\ge T_\beta\}}
{c + \sum_{\ell \in [p]} \ind\{W_\ell \le -T_\beta\}}-
\frac{p\ind\{W_j \le - T_\beta\}}{\sum_{\ell \in [p]} \ind\{W_\ell \le -T_\beta\}}
+\frac{p\ind\{W_j \le - T_\beta\}}{\sum_{\ell \in [p]} \ind\{W_\ell \le -T_\beta\}}
\bigg] \\
& = \EE[\phi_j(\hat t_j;\bm{S}_j)] + \EE\bigg[
\frac{p\ind\{W_j \le -T_\beta\}}{\sum_{\ell \in [p]} \ind\{W_\ell \le -T_\beta\}}
\bigg] \\
& \le \EE\bigg[
\frac{p\ind\{W_j \le -T_\beta\}}{\sum_{\ell \in [p]} \ind\{W_\ell \le -T_\beta\}}
\bigg]. 
\end{align*}
Now suppose instead that $\phi_j(\hat t_j; \bm{S}_j) > 0$. By definition, 
there exists an increasing sequence $t_{j,k} \le \hat t_j$ such that $t_{j,k} \rightarrow \hat t_j$ as $k \rightarrow \infty$ and 
$\phi(t_{j,k};\bm{S}_j) \le 0$. Then 
\begin{align*} 
\EE[e_j] & = \EE\bigg[\frac{p \ind\{\hat{t}_j \cdot W_j> T_\beta\}}
{c + \sum_{\ell \in [p]} \ind\{W_\ell \le -T_\beta\}}\bigg] 
 = \EE\bigg[\lim_{k \rightarrow \infty}\frac{p \ind\{t_{j,k} \cdot W_j\ge  T_\beta\}}
{c + \sum_{\ell \in [p]} \ind\{W_\ell \le -T_\beta\}}\bigg] \\
& \stackrel{\text{(i)}}{=} \lim_{k\rightarrow \infty}\EE\bigg[\frac{p \ind\{t_{j,k} \cdot W_j\ge  T_\beta\}}
{c + \sum_{\ell \in [p]} \ind\{W_\ell \le -T_\beta\}}\bigg] \\
& = \lim_{k\rightarrow \infty}\EE[\phi_j(t_{j,k};\bm{S}_j)]
+ \EE\bigg[\frac{p\ind\{W_j\le - T_\beta\}}
{\sum_{\ell \in [p]} \ind\{W_\ell \le -T_\beta\}}\bigg]\\ 
& \le \EE\bigg[\frac{p\ind\{W_j\le - T_\beta\}}
{\sum_{\ell \in [p]} \ind\{W_\ell \le -T_\beta\}}\bigg],
\end{align*}
where step (i) is by the monotone convergence theorem.

In both cases, there is 
\begin{align*} 
\sum_{j \in \mathcal{H}_0}\EE[e_j] \le  
\sum_{j \in \mathcal{H}_0}\EE\bigg[\frac{\ind\{W_j\le - T_\beta\}}
{\sum_{\ell \in [p]}p \ind\{W_\ell \le -T_\beta\}}\bigg]\le 1,
\end{align*}
verifying that the $e_j$'s are compound e-values. The FDR control 
of the eBH procedure follows directly from~\citet{eBH}.

\subsection{Proof of Theorem~\ref{thm:finite_fdr}}
\label{appendix:proof_finite_fdr}
Roughly speaking, the FDR can be well controlled when $\widehat{\si}_j \le 1$ for 
all $j\in\cH_0$ with high probability. To formalize this idea, for some $\epsilon >0$, we 
decompose the FDR as 
\begin{align*}
\fdr & = \EE\left[ \frac{\sum_{j\in \cH_0} \ind\{W_j \ge T_\alpha\}}{\sum_{j\in[p]} \ind\{W_j \ge T_\alpha\}}\right]\\
& \le \EE\left[ \ind\Big\{\max_{j\in \cH_0} \widehat{\text{SI}}_j \le e^\epsilon \Big\}
\cdot\frac{\sum_{j\in \cH_0} \ind\{W_j \ge T_\alpha\}}
{\sum_{j\in[p]} \ind\{W_j \ge T_\alpha\}}\right] + 
\PP\left(\max_{j\in \cH_0} \widehat{\text{SI}}_j > e^\epsilon\right),
\end{align*}
where the inequality follows because the FDP is upper bounded by one.
By the choice of $T_\alpha$, we further bound the first term above as
\begin{align*}
& \EE\left[\ind\Big\{\max_{j\in \cH_0} \widehat{\text{SI}}_j \le e^\epsilon \Big\} 
\cdot\frac{\sum_{j\in \cH_0} \ind\{W_j \ge T_\alpha\}}{\sum_{j\in [p]} \ind\{W_j \ge  T_\alpha\}}
\right]\\
\le~& \alpha \EE\left[ \ind\Big\{\max_{j\in \cH_0} \widehat{\text{SI}}_j \le e^\epsilon \Big\} 
\frac{\sum_{j\in \cH_0} \ind\{W_j \ge T_\alpha\}}{1 + \sum_{j\in \cH_0} \ind\{W_j \le  -T_\alpha\}}\right]\\
=~& \alpha \sum_{j \in \cH_0} 
\EE\left[\ind\Big\{\max_{\ell \in \cH_0} \widehat{\text{SI}}_\ell \le e^\epsilon \Big\}  \frac{\ind\{W_j \ge T_\alpha\}}{1 + \sum_{\ell \in \cH_0} \ind\{W_{\ell} \le -T_\alpha\}}\right]\\
\le~&\alpha \sum_{j \in \cH_0} 
\EE\left[ \frac{\ind\{\widehat{\text{SI}}_j \le e^\epsilon \} 
\cdot \ind\{W_j \ge T_\alpha\}}{1 + \sum_{\ell \in \cH_0} \ind\{W_{\ell} \le -T_\alpha\}}\right]
\end{align*}
Since the threshold $T_\alpha$ is determined by $(W_1,\ldots,W_p)$, 
we can write $T_\alpha = \mathcal{T}(W_1,\ldots,W_p)$ for some mapping $\mathcal{T}$.
For each $j\in\cH_0$, we further define 
$T_j = \mathcal{T}(W_1,\ldots,|W_j|, \ldots,W_p)$.
On the event $\{W_j \ge T_\alpha\}$, $W_j = |W_j|$ and therefore 
$T_\alpha = T_j$, deterministically. As a result, 
\begin{align*}
\EE\left[\frac{\ind\{W_j \ge T_\alpha\}
\cdot\ind\{\widehat{\text{SI}}_j \le e^\epsilon \} }{1 + \sum_{\ell \in \cH_0}\ind\{W_{\ell} \le -T_\alpha\}}\right]
= \EE\left[\frac{\ind\{W_j \ge T_j\}\cdot
\ind\{\widehat{\text{SI}}_j \le e^\epsilon \}  }{1 + \sum_{\ell \in \cH_0\backslash\{j\}}
\ind\{W_{\ell} \le -T_j\}}\right].
\end{align*}
Recall that, by definition,
\begin{align*}
\widehat{\text{SI}}_j =
\frac{\PP(W_j >0 \,\big|\, |W_j|, W_{\backslash j}\big)}{
\PP(W_j <0 \,\big|\, |W_j|, W_{\backslash j}\big)}. 
\end{align*}
Then by the law of total expectation, 
\begin{align}\label{eq:perj_fdr}
& \EE\left[\frac{\ind\{W_j \ge T_j\}\cdot 
\ind\{\widehat{\text{SI}}_j \le e^\epsilon \}  }
{1 + \sum_{\ell \in \cH_0 \backslash \{j\}}
\ind\{W_{\ell} \le -T_j\}}\right] \notag \\
=~& \EE\left[\EE\left[\frac{\ind\{W_j \ge T_j\} \cdot\ind\{ 
\widehat{\text{SI}}_j \le e^\epsilon \}  }{1 + \sum_{\ell \in \cH_0 \backslash \{j\}}\ind\{W_{\ell} \le -T_j\}} \,\bigg|\,
|W_j|, W_{\backslash j}\right] \right] \notag \\
=~& \EE\left[\frac{\ind\{|W_j| \ge T_j\} \cdot \ind\{\widehat{\text{SI}}_j \le e^\epsilon \}  }{1 + \sum_{\ell \in \cH_0\backslash \{j\}}\ind\{W_{\ell} \le -T_j\}} 
\cdot 
\PP(W_j >0 \,\big|\,
|W_j|, W_{\backslash j} \big) \right],
\end{align}
where the last step is because that $T_j$, $|W_j|$, $W_{\backslash j}$, and 
$\hat{\textnormal{SI}}_j$ are 
deterministic given $|W_j|$ and $W_{\backslash j}$. 
We then have
\begin{align*}    
\eqref{eq:perj_fdr} & \le 
e^\epsilon \cdot  \EE\left[\frac{\ind\{|W_j| \ge T_j\}}
{1 + \sum_{\ell \in \cH_0\backslash\{j\}}\ind\{W_{\ell} \le -T_j\}} 
\cdot 
\PP(W_j < 0 \,\big|\, |W_j|, W_{\backslash j}\big) \right] \\
& = e^\epsilon \cdot  \EE\left[\frac{\ind\{W_j \le -T_j\}}
{\sum_{\ell \in \cH_0}\ind\{W_{\ell} \le -T_j\}} \right]\\
& = e^\epsilon \cdot  \EE\left[\frac{\ind\{W_j \le -T_j\}}
{\sum_{\ell \in \cH_0}\ind\{W_{\ell} \le -T_\ell\}} \right],
\end{align*}
where the last step follows from~\citet[Lemma 6]{barber2020robust}.
Summing over $j \in \cH_0$ and combining everything above, 
we arrive at 
\begin{align*}
\fdr \le e^\epsilon \cdot \alpha + \PP\left(\max_{j \in \cH_0} \hat{\textnormal{SI}}_j > e^\epsilon\right).
\end{align*}

\subsection{Proof of Theorem \ref{thm:fdr}}
\label{appendix:fdr_proof}
We start by defining the following empirical quantities: 
\begin{align*}
&V_-(t) = \frac{1}{p_0}\sum_{j \in \cH_0} \ind\{W_j \le -t\},
~
V_+(t) = \frac{1}{p_0}\sum_{j \in \cH_0} \ind\{W_j \ge t\},\\
& R_-(t) = \frac{1}{p}\sum_{j \in [p]} \ind\{W_j \le -t\},~
R_+(t) = \frac{1}{p}\sum_{j \in [p]} \ind\{W_j \ge t\}, 
\end{align*}
as well as their population counterparts:
\begin{align*} 
F_0(t) = \frac{1}{p_0}\sum_{j\in \cH_0} \PP(W_j \ge t),
~
G_-(t) = \frac{1}{p}\sum_{j \in [p]} \PP(W_j \le  -t),~
G_+(t) = \frac{1}{p}\sum_{j \in [p]} \PP(W_j \ge t).
\end{align*}
By symmetry, we have that $F_0(t) = \EE[V_-(t)] = \EE[V_+(t)]$. 
We also let 
\begin{align*}
& \fdp(t) = \frac{p_0 V_+(t)}{\max(pR_+(t),1)},
~ 
\widehat{\fdp}(t) = \frac{pR_-(t)}{\max(pR_+(t),1)},
~
\overline{\fdp}(t) = \frac{\pi_0F_0(t)}{G_+(t)},
~
\widetilde{\fdp}(t) = \frac{G_-(t)}{G_+(t)}
\end{align*}
which corresponds to the true, estimated, and their asymptotic counterparts of the FDP.

To connect the empirical quantities with their oracle counterparts, 
we leverage the following lemma adapted from~\citet{GM}. Note that 
the condition and proof in~\citet{GM} only apply to the proof of 
$V_-(t)$'s uniform convergence. We nevertheless modify their proof and 
provide the details in Section~\ref{sec:proof_concentration}.
\begin{lemma}[Lemma 8 of~\citet{GM}]
\label{lem:concentration}
Under the same assumption of Theorem~\ref{thm:fdr}, and 
suppose that $F_0(t)$ is a continuous function. Then
as $p_0,p_1 \rightarrow \infty$, 
\begin{align*}
& \sup_t |V_-(t) - F_0(t)| \covp 0, 
~
\sup_t |V_+(t) - F_0(t)| \covp  0,\\
~
& \sup_t |R_-(t) - G_-(t)| \covp 0,
~
\sup_t |R_+(t) - G_+(t)| \covp 0.
\end{align*}
\end{lemma}
By assumption, for any sufficiently small $\varepsilon > 0$, 
there exists a threshold $t_\eps$ such that 
\begin{align*}
\widetilde{\fdp}(t_\eps) \le \alpha - \varepsilon
\text{ and } G_+(t_\eps)>0.
\end{align*}
Also, since $\hat{\pi}_0 :=p_0/p \rightarrow \pi_0$ as $p \rightarrow \infty$, 
there exists sufficiently large $m_0 \in \mathbb{N}_+$ such that 
for any $p_0,p_1 \ge m_0$, $|p_0 / p - \pi_0| \le \eps$.

Using Lemma~\ref{lem:concentration}, we have for the fixed $t_\eps$, 
$R_-(t_\eps) \covp G_-(t_\eps)$, and $R_+(t_\eps) \covp G_+(t_\eps)$. 
By the continuous mapping theorem,
\begin{align*}
\widehat{\fdp}(t_\eps) = \frac{R_-(t_\eps)}{R_+(t_\eps)}
\covp \frac{G_-(t_\eps)}{G_+(t_\eps)} = \widetilde{\fdp}(t_\eps).
\end{align*}
Therefore, for any $\delta \in (0,1)$, 
there exists $m_1 \in \mathbb{N}_+$ large enough such that for any 
$p_0,p_1 > m_1$, 
\begin{align*}
\PP\Big(\big|\widehat{\fdp}(t_\eps) - \widetilde{\fdp}(t_\eps)\big| \le \eps/2 \Big) \ge 
1-\delta 
\Rightarrow 
\PP\big(\widehat{\fdp}(t_\eps) \le \alpha -\eps/2 \big) \ge 1-\delta.
\end{align*}
On the event $\{\widehat{\fdp}(t_\eps) \le \alpha - \eps/2\}$, there is 
$T_\alpha \le t_\eps$ (recall the definition of $T_\alpha$). 
By monotonicity, 
$R_+(T_\alpha) \ge R(t_\eps)$ when $T_\alpha \le t_\eps$. 
Meanwhile, since $|R_+(t_\eps) - G_+(t_\eps)| \covp 0$, 
we can find $m_2 \in \mathbb{N}_+$ large enough such that when $p_0,p_1 \ge m_2$, 
with probability at least $1-\delta$, 
\begin{align*}
 |R_+(t_\eps)- G_+(t_\eps)| \le G_+(t_\eps)/2 \Rightarrow R_+(t_\eps) \ge G_+(t_\eps)/2 >0.
\end{align*} 
Next, on the event $\{T_\alpha \le t_\eps, R_+(t_\eps) \ge G_+(t_\eps)/2\}$, 
\begin{align*}
& \big|\widehat{\fdp}(T_\alpha) - \widetilde{\fdp}(T_\alpha)\big| \\
\le~& 
\bigg|\frac{R_-(T_\alpha)}{R_+(T_\alpha)} - \frac{G_-(T_\alpha)}{R_+(T_\alpha)}\bigg|
+ \bigg|\frac{ G_-(T_\alpha)}{R_+(T_\alpha)} - \frac{G_-(T_\alpha)}{G_+(T_\alpha)}\bigg|\\
\le~& \frac{2}{G_+(t_\eps)}\big|R_-(T_\alpha) - G_-(T_\alpha)\big|
+ G_-(T_\alpha) \bigg|\frac{1}{R_+(T_\alpha)} - \frac{1}{G_+(T_\alpha)}\bigg|\\ 
=~& \frac{2}{G_+(t_\eps)}\big|R_-(T_\alpha) - G_-(T_\alpha)\big|
+ \frac{G_-(T_\alpha)}{R_+(T_\alpha)G_+(T_\alpha)} 
\big|R_+(T_\alpha) - G_+(T_\alpha)\big|\\
\le~& \frac{2}{G_+(t_\eps)^2} \Big\{
\sup_{t \le t_\eps} |R_-(t) - G_-(t)| + 
\sup_{t \le t_\eps} |R_+(t) - G_+(t)|\Big\}.
\end{align*}
Similarly,  on the same event, with $p_0,p_1 \ge m_0$, we have  
\begin{align*}
& \Big|\fdp(T_\alpha) - \afdp(T_\alpha) \Big|\\
\le  ~&
\frac{V_+(T_\alpha)}{R_+(T_\alpha)} |\hat \pi_0 - \pi_0|
+ \bigg|\frac{\pi_0 V_+(T_\alpha)}{R_+(T_\alpha)} - \frac{\pi_0 F_0(T_\alpha)}{R_+(T_\alpha)}\bigg|
+ \bigg|\frac{\pi_0F_0(T_\alpha)}{R_+(T_\alpha)} - \frac{\pi_0F_0(T_\alpha)}{G_+(T_\alpha)}\bigg|\\
\le~& \eps + \frac{2}{G_+(t_\eps)}|V_+(T_\alpha) - F_0(T_\alpha)|
+ F_0(T_\alpha) \bigg|\frac{1}{R_+(T_\alpha)} - \frac{1}{G_+(T_\alpha)}\bigg|\\ 
=~ & \eps + \frac{2}{G_+(t_\eps)}|V_+(T_\alpha) - F_0(T_\alpha)|
+ \frac{F_0(T_\alpha)}{R_+(T_\alpha)G_+(T_\alpha)} 
\big|R_+(T_\alpha) - G_+(T_\alpha)\big|\\
\le~& \eps+ \frac{2}{G_+(t_\eps)} \Big\{
\sup_{t \le t_\eps} |V_+(t) - F_0(t)| +
 \sup_{t \le t_\eps} |R_+(t) - G_+(t)|\Big\}
\end{align*}
Again by Lemma~\ref{lem:concentration}, 
there exists $m_3 \in \mathbb{N}_+$ such that 
when $p_0,p_1 \ge m_3$, with probability at least $1-\delta$,  
\begin{align}
\label{eq:supconv_event}
& \sup_t |V_+(t) - F_0(t)| \le G_+(\eps)\eps,~
\sup_t |V_-(t) - F_0(t)| \le G_+(\eps)\eps,\\
& \sup_t |R_+(t) - G_+(t)| \le G_+(\eps)\eps, ~
 \sup_t |R_-(t) - G_-(t)| \le G_+(\eps)\eps, ~
\end{align}
Let $A$ denote the intersection of $\{T_\alpha \le t_\eps, R(t_\eps) \ge G(t_\eps)\}$ and 
the event in Equation~\eqref{eq:supconv_event}. 
Taking the union bound, we have that $\PP(A^c) \le 3\delta$.
Then, for $p_0,p_1 \ge \max(m_0,m_1,m_2,m_3)$,
\begin{align*}
& \PP\big(\fdp(T_\alpha) - \widehat{\fdp}(T_\alpha) \ge 11\eps \big)\\
\le~&  \PP\big(\fdp(T_\alpha) - \widehat{\fdp}(T_\alpha)\ge 11\eps, A \big) + 3\delta\\
\stackrel{\textnormal{(a)}}{\le}~& \PP\big({\fdp}(T_\alpha) - \afdp(T_\alpha) + 
\widetilde{\fdp}(T_\alpha) - \widehat{\fdp}(T_\alpha) \ge 11\eps, A\big) + 3\delta\\
\stackrel{\textnormal{(b)}}{\le}~& 3\delta.
\end{align*}
Above, step (a) is because $\overline{\fdp}(T_\alpha) \le \widetilde{\fdp}(T_\alpha)$
and step (b) is because on the event $A$, 
\begin{align*}
\big|\widehat{\fdp}(T_\alpha) -  \widetilde{\fdp}(T_\alpha)\big| + 
\big|\afdp(T_\alpha) - \fdp(T_\alpha)\big| \le 10\eps.
\end{align*}
We then proceed to upperbound the FDR. When $p_0,p_1 \ge \max(m_0,m_1,m_2,m_3)$, 
\begin{align*}
\textnormal{FDR} & = \EE\big[\fdp(T_\alpha)\big]\\
& \le \EE\big[\fdp(T_\alpha) \cdot \ind\{\fdp(T_\alpha) - \widehat{\fdp}(T_\alpha)| < 11\eps\}\big]
+ 3\delta\\
& \le \EE\big[\widehat{\fdp}(T_\alpha)\big] + 11\eps + 3\delta \\ 
& \le \alpha + 11\eps + 3\delta,
\end{align*}
where the last inequality follows from the definition of $T_\alpha$. 
Since $\eps$ and $\delta$ are arbitrary, we conclude the proof.

\subsection{Proof of Lemma~\ref{lem:concentration}}
\label{sec:proof_concentration}
For any $\eps \in (0,1)$, we let $N_\eps = \lceil 2 / \eps \rceil$.
Since $F_0(t)$ is continuous in $t$, we can find  
$-\infty = a_0 \le a_1 \cdots \le  a_{N_\eps} = \infty$ such that 
$F_0(a_k) - F_0(a_{k-1})\le \eps/2$, $\forall k \in [N_\eps]$.
Then,
\begin{align*}
& \PP\big(\sup_{t} |V_-(t) - F_0(t)| \ge \eps\big)
\le \PP\Big(\bigcup_{k \in [N_\eps]} \sup_{t \in (a_{k-1},a_k]}|V_-(t) - F_0(t)| \ge \eps\Big)\\ 
\le ~&\sum_{k \in [N_\eps]} 
\PP\Big(\sup_{t \in (a_{k-1},a_k]}|V_-(t) - F_0(t)| \ge \eps\Big)
\le \sum_{k \in [N_\eps]} 
\PP\big(|V_-(a_{k-1}) - F_0(a_{k-1})| \ge \eps/2\big),
\end{align*}
where the last step follows from the choice of $a_k$'s.
Invoking Chebyshev's inequality, we have 
\begin{align*}
\sum_{k \in [N_\eps]}\PP\big(|V_-(a_{k-1}) - F_0(a_{k-1})| \ge \eps/2\big)
& \le \sum_{k \in[N_\eps]}\frac{\sum_{i,j \in \cH_0} \textnormal{Cov}(\ind\{W_i \le -a_{k-1}\}, 
\ind\{W_j \le -a_{k-1}\}) }{p_0^2 \eps^2 / 4} \\
& \le \sum_{k \in [N_\eps]}\frac{c\cdot p^\kappa}{p_0^2 \eps^2/4}
= \frac{8c} {\pi_0^2 p^{2-\kappa}\eps^3}, 
\end{align*}
which goes to zero as $p\rightarrow \infty$ since $\kappa \in(0,2)$.
The proof for the convergence of $V_+$ is exactly the same. 

We now focus on $R_-(t)$. Again by the continuity of $G_-(t)$, we can find
$-\infty = b_0 \le b_1 \cdots \le  b_{N_\eps} = \infty$ such that 
$G_-(b_k) - G_-(b_{k-1})\le \eps/2$, $\forall k \in [N_\eps]$.
Then,
\begin{align*}
& \PP\big(\sup_{t} |R_-(t) - G_-(t)| \ge \eps\big)
\le \PP\Big(\bigcup_{k \in [N_\eps]} \sup_{t \in (b_{k-1},b_k]}|R_-(t) - G_-(t)| \ge \eps\Big)\\ 
\le ~&\sum_{k \in [N_\eps]} 
\PP\Big(\sup_{t \in (b_{k-1},b_k]}|R_-(t) - G_-(t)| \ge \eps\Big)
\le \sum_{k \in [N_\eps]} 
\PP\big(|R_-(b_{k-1}) - G_-(b_{k-1})| \ge \eps/2\big),
\end{align*}
where the last step follows from the choice of $b_k$'s.
Invoking Chebyshev's inequality, we have 
\begin{align*}
\sum_{k\in [N_\eps]}\PP\Big(|R_-(b_{k-1}) - G_-(b_{k-1})| \ge \eps/2\Big)
& \le \sum_{k\in [N_\eps]}\frac{\sum_{i,j \in [p]} \textnormal{Cov}(\ind\{W_i \le -b_{k-1}\}, 
\ind\{W_j \le -b_{k-1}\}) }{p^2 \eps^2 / 4} \\
& \le \sum_{k\in [N_\eps]}\frac{c\cdot p^\kappa}{p^2 \eps^2/4}
= \frac{8c}{p^{2-\kappa}\eps^3}, 
\end{align*}
which goes to zero as $p\rightarrow \infty$ since $\kappa \in(0,2)$.
The proof for the convergence of $R_+$ is exactly the same.

\section{Additional simulation details}
\label{appendix:sim}
\subsection{Implementation details of the Gaussian mirror}

Our implementation of the GM procedure in our $p<n$ setting matches the algorithm that was theoretically proven by \cite[Theorem 4]{GM} to exhibit asymptotic FDR control. Specifically, for $j\in [p]$, GM constructs $\bm z_j \sim N(0, c_j \bm I_n)$, where the marginal variance $c_j$ is defined as
\[ c_j = \sqrt{\frac{\bm{x}_j^{\top} \left( \bm I_n - \bm{X}_{-j} (\bm{X}_{-j}^{\top} \bm{X}_{-j})^{-1} \bm{X}_{-j}^{\top} \right) \bm{x}_j}{\bm{z}_j^{\top} \left( \bm I_n - \bm{X}_{-j} (\bm{X}_{-j}^{\top} \bm{X}_{-j})^{-1} \bm{X}_{-j}^{\top} \right) \bm{z}_j}}.
 \]
GM then regresses $\bm y$ onto $[\bm x_j + \bm z_j, \bm x_j - \bm z_j, \bm X_{-j}]$, using OLS or regularized regression. Letting $\widehat\beta_j^+$ and $\widehat\beta_j^-$ be the coefficients for $\bm x_j + \bm z_j$ and $\bm x_j - \bm z_j$, respectively, GM 
computes the test statistic $W_j = \lvert \widehat\beta_j^+ + \widehat\beta_j^- \rvert - \lvert \widehat\beta_j^+ - \widehat\beta_j^- \rvert$. Finally, it constructs the rejection threshold $T_\alpha$ and returns $\widehat S$ in the same manner as BC and OATK using Equation \eqref{eq:W_thresh}. As noted in \citet{GM}, the Gaussian mirror $\bm z_j$ can be viewed as a type of knockoff for $\bm x_j$ since the procedure is equivalent to regressing $\bm y$ onto $[\bm X, \bm z_j]$ and then taking $W_j$ to be the difference in the magnitude of the regression coefficients for $\bm x_j$ and $\bm z_j$. This formulation achieves asymptotic FDR control when using OLS under a mild correlation assumption on $\bm X$. We use Lasso regression in our implementation, as we found it yields higher power than its OLS and ridge versions while largely maintaining FDR control in our numerical examples. 

This implementation differs from what was used in their numerical studies and coded in their accompanying software \citep{gm_code}, primarily due to a pre-screening step that is necessary for $p>n$. They initially pre-screen for promising variables by conducting Lasso regression on the original data. Let $R$ be the set of indices with nonzero coefficients in the pre-screening regression and $\widehat{ \bm \beta}_{R}$ be their coefficients. They obtain $\bm X_R$, which consists of only the columns of $\bm X$ in $R$. Then, they construct Gaussian mirror variables for all $p$ variables, but they select $c_j$ only with respect to $\bm X_R$, such that
\begin{equation}
    c_j = \sqrt{\frac{\bm{x}_j^{\top} \left( \bm I_n - \bm{X}_{R\backslash j} (\bm{X}_{R\backslash j}^{\top} \bm{X}_{R\backslash j})^{-1} \bm{X}_{R\backslash j}^{\top} \right) \bm{x}_j}{\bm{z}_j^{\top} \left( \bm I_n - \bm{X}_{R\backslash j} (\bm{X}_{R\backslash j}^{\top} \bm{X}_{R\backslash j})^{-1} \bm{X}_{R\backslash j}^{\top} \right) \bm{z}_j}}, \label{eq:c_screen}
\end{equation}
where $\bm{X}_{R\backslash j}$ is $\bm X_R$ if $j\notin R$ and $\bm X_R$ with the $\bm x_j$ column removed if $j\in R$. The software package of GM \citep{gm_code} uses this pre-screened formulation for both $p < n$ and $p > n$ cases. We found the Lasso pre-screening step to boost the power and FDR control in both the GM and OATK procedures, although OATK with pre-screening was still more powerful than GM with pre-screening.

In our numerical implementations in Section \ref{sec:numerical}, we implement both GM and OATK without the pre-screening step for multiple reasons. First, pre-screening was primarily intended to handle the $p > n$ case, whereas our focus is on $p < n$, for which pre-screening is not needed. Second, it allows the more innate aspects of the two methods to be compared without conflating the effects of pre-screening. Third, the pre-screening coded implementation in Xing, et al. (2023b) lacks a firm theoretical basis. Regarding the last point, while the coded implementation follows to some extent the high-dimensional ($p>n$) GM algorithm for which \citet[Algorithm 2, Theorem 5]{GM} proved asymptotic FDR control, there are fundamental differences that render the theoretical results inapplicable. 
First, Algorithm 2 of \citet{GM} generates $\bm z_j$ only for $j\in R$, i.e., only the variables that survive the pre-screening step, and the screened-out variables are completely ignored in the subsequent GM regressions and considered as null variables by default. In contrast, the implemented algorithm generates $\bm z_j$'s for all variables, including the screened-out $j$'s, which may be included in $\widehat S$ depending on $W_j$ and $T_\alpha$. We found empirically that ignoring all screened-out variables resulted in much poorer FDR control than when knockoffs were generated for them, as was done in the coded implementation. Moreover, $c_j$ in Algorithm 2 of \citet{GM} is computed as
\begin{equation}
    c_j = \sqrt{\frac{\bm{x}_j^{\top} \left( \bm I_n - \bm{X}_{R\backslash j} (\bm{X}_{R\backslash j}^{\top} \bm{X}_{R\backslash j})^{-1} \bm{X}_{R\backslash j}^{\top} \right) \bm{x}_j}{\widetilde{\bm{z}}_j^{\top} \left( \bm I_n - \bm{X}_{R\backslash j} (\bm{X}_{R\backslash j}^{\top} \bm{X}_{R\backslash j})^{-1} \bm{X}_{R\backslash j}^{\top} \right) \widetilde{\bm{z}}_j}}, 
\end{equation}
where $\widetilde{\bm{z}}_j$ is the projection of $\bm z_j$ onto the column space of $\bm X_R$, which differs substantially from \eqref{eq:c_screen} used in the coded implementation.

Additionally, the construction of $W_j$ differs between the two algorithms. The coded implementation constructs $W_j = \lvert \widehat\beta_j^+ + \widehat\beta_j^- \rvert - \lvert \widehat\beta_j^+ - \widehat\beta_j^- \rvert$, which is the same as in the low-dimensional algorithm without the pre-screening step. In contrast, Algorithm 2 defines
\begin{equation}
    W_j = \left| \sigma \Phi^{-1} \left (F^{[\mathcal{V}_1^-(r), \mathcal{V}_1^+(r)]}_{0, \sigma^2} \left( \hat{\beta}_j^+ + \hat{\beta}_j^- \right) \right) \right| - \left| \sigma \Phi^{-1} \left( F^{[\mathcal{V}_0^-(r), \mathcal{V}_0^+(r)]}_{0, \sigma^2} \left( \hat{\beta}_j^+ - \hat{\beta}_j^- \right) \right) \right|, \label{eq:gm_screen}
\end{equation}
where $\sigma$ is the standard deviation of the noise term of the linear model \eqref{eq:linear_model} (which must be estimated), $\Phi(\cdot)$ is the CDF of the standard normal distribution, $F_{\mu, \sigma^2}^{[a, b]}$ is the CDF of a $N(\mu, \sigma^2)$ random variable truncated to the interval $[a, b]$, and $\mathcal{V}_1^-(r), \mathcal{V}_1^+(r), \mathcal{V}_0^-(r), \mathcal{V}_0^+(r)$ are parameters calculated from $\bm X_R$, $\bm y$,  and $\widehat{ \bm \beta}_{R}$.
Defining $W_j$ as in \eqref{eq:gm_screen} is necessary to prove asymptotic FDR control because it results in the symmetry of $W_j$ for null $j$'s by correcting for the post-selection bias caused by the Lasso regression of the pre-screening step, whereas the version of $W_j$ implemented in the code does not.

\subsection{Additional numerical results for Gaussian $\bm X$}
\label{appendix:gaussian}

\begin{figure}
    \centering
    \includegraphics[width=\textwidth]{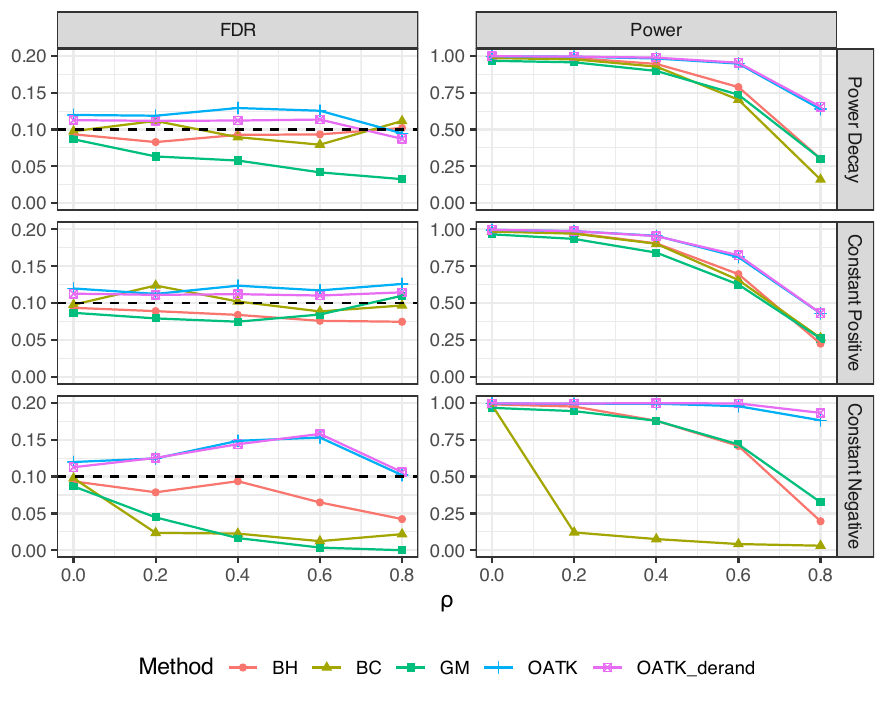}
    \vspace*{-20mm}
    \caption{FDR and power as functions of covariance parameter $\rho$ in the $p=300$ and $n=1000$ Gaussian $\bm X$ study with fixed signal amplitude $A=6$.}
    \label{fig:gaussian-main-small-rho}
\end{figure}

This section examines additional numerical results for the Gaussian $\bm X$ study of Section \ref{subsec:gaussian_numerical}. Fig. \ref{fig:gaussian-main-small-rho} shows the effect of the covariance parameter $\rho$ on the FDR and power on the smaller $n=1000, p=300$ example with fixed amplitude $A=6$. Increasing $\rho$, which increases the correlation among covariates in all three covariance structures, reduces power across all variable selection procedures. However, OATK shows the slowest decay in power, and increasing $\rho$ does not change OATK's uniformly higher power compared to other algorithms. The FDR control of OATK is consistently reasonable across $\rho$ in the power decay and constant positive examples, never exceeding 0.13. In the constant negative case, OATK achieves the worst FDR inflation at $0.15$ when $\rho=0.6$, but we view such inflation as relatively innocuous, given OATK's superior power and the replicate-to-replicate variability, 
as discussed in Section \ref{subsec:gaussian_numerical}.

\subsection{Fast implementation of conditionally calibrated OATK}
\label{appendix:calibration}

We run the conditional calibration procedure detailed in Section \ref{sec:ext} on $\widetilde \cS = \bigl( R^{BH(4\alpha)} \cap \cS^{p}\bigr) \;\cup\; \cS^{Kn}\cup \widehat S$, where $R^{BH(4\alpha)}$ is the rejection set of BH with FDR level $4\alpha$, $\widehat S$ is the rejection set of the basic OATK, and 
\begin{align}
\cS^{p} &= \{\, j : p_j \leq  \alpha/2 \,\}, \\
\cS^{Kn} &= \{\, j : |W_j| \ge w_{\,p - |R^{BH(4\alpha)}\cap \cS^{p}|} \wedge w_{T_\alpha} \,\}.
\end{align}
Here, $p_j$ denotes the p-value corresponding to the standard two-sided t-statistic from OLS and $w_1 < \cdots < w_p$ denote the order statistics of $|W_1|, \cdots, |W_p|.$ In other words, we run the calibration procedure only on variables with relatively small p-value, a larger OATK test statistic, or in the rejection set of OATK.